\documentclass[11pt]{article}
\newtheorem{Def}{Def.}[section]
\newtheorem{Thm}[Def]{Theorem}

\newtheorem{Lemma}[Def]{Lemma}

\newcommand{\Proof}{{\em{Proof: }}}
\newcommand{\QED}{\ \hfill $\FBox$ \\[1em]}
\newcommand{\Equ}[1]{\begin{equation} \label{eq:#1}}
\newcommand{\EndEqu}{\end{equation}}

\newcommand{\Pdd}{\mbox{$\partial$ \hspace{-1.2 em} $/$}}

\newcommand{\spc}{\;\;\;\;\;\;\;\;\;\;}

\newcommand{\Aslsh}{\mbox{ $\!\!A$ \hspace{-1.2 em} $/$}}
\newcommand{\slsh}{\mbox{ \hspace{-1.1 em} $/$}}

\newcommand{\R}{\mbox{\rm I \hspace{-.8 em} R}}
\newcommand{\1}{\mbox{\rm 1 \hspace{-1.05 em} 1}}
\newcommand{\Z}{\mbox{\rm \bf Z}}

\newcommand{\FBox}{\rule{2mm}{2.25mm}}
\newcommand{\OBox}{\raisebox{.6ex}{\fbox{}}\,}

\title{Light-Cone Expansion of the Dirac Sea to First Order in the 
External Potential}
\author{Felix Finster\thanks{Supported by the Deutsche 
Forschungsgemeinschaft, Bonn.}\\ Department of Mathematics, Harvard University}
\date{July 1997 / September 1998}

\begin{document}
\maketitle
\begin{abstract}
The perturbation of the Dirac sea to first order in the external 
potential is calculated in an expansion around the light cone. It is 
shown that the perturbation consists of a causal contribution, which 
describes the singular behavior of the Dirac sea on the light cone 
and contains bounded line integrals over the potential and its partial 
derivatives, and a noncausal contribution, which is a smooth function.
As a preparatory step, we construct a formal solution of the 
inhomogeneous Klein-Gordon equation in terms of an infinite series of
line integrals.

More generally, the method presented can be used for an explicit 
analysis of Feynman diagrams of the Dirac, Klein-Gordon, and wave equations in 
position space.
\end{abstract}

\section{Introduction}
\setcounter{equation}{0}
In relativistic quantum mechanics, the problem of the unphysical 
negative-energy solutions of the Dirac equation is solved by the 
conception that all negative-energy states are occupied in the vacuum 
forming the so-called Dirac sea. In \cite{F1}, the Dirac sea was 
constructed for the Dirac equation with general interaction in terms 
of a formal power series in the external potential.
In the present paper, we turn our attention to a single Feynman diagram
of this perturbation expansion. 
More precisely, we will analyze the contribution to first order in the 
potential and derive explicit formulas for the 
Dirac sea in position space.
Since this analysis does not require a detailed knowledge of the
perturbation expansion for the Dirac sea, we can make this paper
self-consistent by giving a brief introduction to the mathematical
problem.

In the vacuum, the Dirac sea is characterized by the integral over the 
lower mass shell
\begin{equation}
        P(x,y) \;=\; \int \frac{d^4p}{(2 \pi)^4} \: (p \slsh + m) \:
        \delta(p^2-m^2) \: \Theta(-p^0) \; e^{-ip(x-y)}
        \label{0}
\end{equation}
($\Theta$ is the Heavyside function $\Theta(x)=1$ for $x \geq 0$ and 
$\Theta(x)=0$ otherwise);
$P(x,y)$ is a tempered distribution that solves the free Dirac 
equation $(i \Pdd_x - m) \: P(x,y) \:=\: 0$.
In the case with interaction, the Dirac sea is accordingly
described by a tempered distribution $\tilde{P}(x,y)$ being a
solution of the Dirac equation
\begin{equation}
        (i \Pdd_x + {\cal{B}}(x) - m) \: \tilde{P}(x,y) \;=\; 0 \;\;\; ,
        \label{1}
\end{equation}
where ${\cal{B}}$ is composed of the classical bosonic potentials. We
assume ${\cal{B}}$ to be a $4 \times 4$ matrix potential satisfying the 
condition $\gamma^0 {\cal{B}}(x)^\dagger \gamma^0={\cal{B}}(x)$ (`$^\dagger$'
denotes the transposed, complex conjugated matrix). We can thus decompose it 
in the form
\begin{equation}
        {\cal{B}} \;=\; e \Aslsh \:+\: e \gamma^5 \:B \!\slsh \:+\: \Phi
\:+\: i \gamma^5 \Xi \:+\: \sigma_{jk} \: H^{jk}
        \label{00}
\end{equation}
with the electromagnetic potential $A_j$, an axial potential $B_j$, scalar 
and pseudoscalar potentials $\Phi$ and $\Xi$, and a bilinear potential
$H^{jk}$ (see e.g.\ \cite{T} for a discussion of these potentials). 
In Appendix \ref{app_b}, it is shown how the results can be extended to 
an external gravitational field.

The Dirac equation (\ref{1}) can be solved by a perturbation expansion.
To first order in ${\cal{B}}$, one has
\[ \tilde{P}(x,y) \;=\; P(x,y) \:+\: \Delta P(x,y) \:+\: 
{\cal{O}}({\cal{B}}^2) \;\;\; , \]
where $\Delta P$ satisfies the inhomogeneous Dirac equation
\begin{equation}
        (i \Pdd_x - m) \: \Delta P(x,y) \;=\; - {\cal{B}}(x) \: P(x,y) \;\;\; .
        \label{2}
\end{equation}
The factor $(i \Pdd_x - m)$ can be inverted with a Green's 
function: We choose as Green's function the sum of the retarded and advanced 
Green's functions,
\begin{equation}
        s(x,y) \;=\; \frac{1}{2} \: \lim_{0<\varepsilon \rightarrow 0}
        \sum_{\pm} \int \frac{d^4p}{(2 \pi)^4} \: \frac{p \slsh + m}{p^2 - 
        m^2 \pm i \varepsilon p^0} \; e^{-ip(x-y)} \;\;\; .
        \label{2a}
\end{equation}
According to its definition, $s$ satisfies the equation
\begin{equation}
        (i \Pdd_x - m) \: s(x,y) \;=\; \delta^4(x-y) \;\;\; .
        \label{3}
\end{equation}
As a consequence, the integral
\begin{equation}
        \Delta P(x,y) \;:=\; - \int d^4z \: \left( s(x,z) \:{\cal{B}}(z)\: P(z,y) \:+\:
        P(x,z) \:{\cal{B}}(z)\: s(z,y) \right)
        \label{4}
\end{equation}
is a solution of (\ref{2}).

Clearly, (\ref{4}) is not the only solution of the inhomogeneous 
Dirac equation (\ref{2}). For example, we could have worked with the advanced
or retarded Green's function instead of (\ref{2a}) or could 
have omitted the second summand in (\ref{4}). The special form of
our solution follows from the causality principle for the 
Dirac sea which was introduced and discussed in \cite{F1}.
We will not repeat these considerations here, and simply take (\ref{4}) 
as an ad hoc formula for the perturbation of the Dirac sea. The reader 
who feels uncomfortable with this procedure is either referred to 
\cite{F1} or can, in a simplified argument, explain the special 
form (\ref{4}) from the ``Hermiticity condition''
\[ \Delta P(x,y)^\dagger \;=\; \gamma^0 \:\Delta P(y,x) \:\gamma^0 \;\;\; , \]
which seems quite natural to impose.

In the language of Feynman diagrams, (\ref{4}) is a first order tree diagram.
In comparison to diagrams of higher order or to loop 
diagrams, this is a very simple diagram, and it might seem 
unnecessary to study the diagram further. Unfortunately, (\ref{4})
gives no information on what $\Delta P$ explicitly looks like in position space.
We are especially interested in the behavior of $\Delta P(x,y)$ 
in a neighborhood of the light cone $(y-x)^2 \equiv (y-x)_j (y-x)^j=0$.
\begin{Def}
\label{def1}
A tempered distribution $A(x,y)$ is of the order 
${\cal{O}}((y-x)^{2p})$, $p \in \Z$, if the product
\[ (y-x)^{-2p} \: A(x,y) \]
is a regular distribution (i.e.\ a locally integrable function).
It has the {\bf{light-cone expansion}}
\begin{equation}
        A(x,y) \;=\; \sum_{j=g}^{\infty} A^{[j]}(x,y)
        \label{6a}
\end{equation}
if the distributions $A^{[j]}(x,y)$ are of the order ${\cal{O}}((y-x)^{2j})$
and if $A$ is approximated by the partial sums in the way that
\begin{equation}
        A(x,y) \;-\; \sum_{j=g}^p A^{[j]}(x,y) \;\;\;
        {\mbox{is of the order}} \;\;\; {\cal{O}}((y-x)^{2p+2})
        \label{6b}
\end{equation}
for all $p \geq g$.
\end{Def}
The first summand $A^{[g]}(x,y)$ gives the leading order of $A(x,y)$ 
on the light cone. If $A$ is singular on the light cone, $g$ will be 
negative. Notice that the $A^{[j]}$ are determined only up to 
contributions of higher order ${\cal{O}}((y-x)^{2j+2})$, but this will 
not lead to any problems in the following.

We point out that we do not study the convergence of the sum 
(\ref{6a}); we only make a statement on the approximation of $A$ by 
the finite partial sums.
The reason why questions of convergence are excluded is that the 
distributions $A^{[j]}$ will typically involve partial derivatives of order
$2j$ of the potential ${\cal{B}}$, and we can thus expect 
convergence only if ${\cal{B}}$ is analytic (for nonanalytic functions, the 
partial derivatives may increase arbitrarily fast in the order of the 
derivative, which makes convergence impossible). Analyticity of the potential,
however, is too strong a condition for physical applications;
we can assume only that ${\cal{B}}$ is 
smooth (the reason why analytic functions are too restrictive is 
that they are completely determined from their behavior in a small 
open set, which contradicts causality).
Thus, the infinite sum in (\ref{6a}) is merely a convenient notation
for the approximation by the partial sums (\ref{6b}).
Despite this formal character of the sum, the light-cone expansion
completely describes the behavior of $A(x,y)$ near the light cone.
This situation can be seen in analogy to writing down the
Taylor expansion for a smooth, nonanalytic function.
Although the Taylor series does not converge in general, the
Taylor polynomials give local approximations of the function.

Our aim is to derive explicit formulas for the light-cone expansion of
$\Delta P(x,y)$.

\section{Discussion of the Method}
\setcounter{equation}{0}
Before performing the light-cone expansion, we briefly discuss the 
basic problem and describe the possible methods for calculating
$\Delta P(x,y)$.

At first sight, our problem seems quite complicated because of the 
Dirac matrices in $s$, in $P$, and in the potential (\ref{00}). Actually, 
this is not the difficult point; we can reduce to a scalar problem by pulling 
all Dirac matrices out of the integral (\ref{4}) as follows. We have
\begin{eqnarray}
        P(x,y) & = & (i \Pdd_x + m) \: T_{m^2}(x,y) \;=\; (-i \Pdd_y + m) \: 
        T_{m^2}(x,y)
        \label{9a} \\
        s(x,y) & = & (i \Pdd_x + m) \: S_{m^2}(x,y) \;=\; (-i \Pdd_y + m) \:
        S_{m^2}(x,y) \;\;\; , \nonumber
\end{eqnarray}
where $T_{m^2}$ and $S_{m^2}$ denote the negative-energy eigenspace 
and the Green's function of the Klein-Gordon operator, respectively:
\begin{eqnarray}
        T_{m^2}(x,y) & = & \int \frac{d^4p}{(2 \pi)^4} \: \delta(p^2-m^2) \:
        \Theta(-p^0) \; e^{-ip(x-y)}
        \label{8a} \\
        S_{m^2}(x,y) & = & \frac{1}{2} \:\lim_{0<\varepsilon \rightarrow 0}
        \sum_{\pm} \int \frac{d^4p}{(2 \pi)^4} \: \frac{1}{p^2-m^2\pm i 
        \varepsilon p^0} \; e^{-ip(x-y)} \;\;\; .
        \label{8b}
\end{eqnarray}
Using the short notation $(\gamma^a)_{a=1,\ldots,16}$ for the bases
$\1, i\gamma^5, \gamma^j, \gamma^5 \gamma^j, \sigma^{jk}$ of the Dirac 
matrices, we can thus rewrite (\ref{4}) in the form
\begin{equation}
        \Delta P(x,y) \;=\; \sum_{a=1}^{16} (i \Pdd_x + m) \:\gamma^a \:
        (-i \Pdd_y + m) \; \Delta T_{m^2}[{\cal{B}}_a](x,y)
        \label{6}
\end{equation}
with
\begin{eqnarray}
\lefteqn{ \Delta T_{m^2}[V](x,y) } \nonumber \\
&=& -\int d^4z \: \left(S_{m^2}(x,z) \:V(z)\: T_{m^2}(z,y) \:+\:
        T_{m^2}(x,z) \:V(z)\: S_{m^2}(z,y) \right) \;\;\;\; .
        \label{7}
\end{eqnarray}
The scalar distribution $\Delta T_{m^2}[V](x,y)$ is a solution of the
inhomogeneous Klein-Gordon equation
\begin{equation}
        (-\OBox_x - m^2) \: \Delta T_{m^2}(x,y) \;=\; -V(x) \: T_{m^2}(x,y)
\;\;\; , \label{8}
\end{equation}
as is immediately verified.
Once we have derived the light-cone expansion for $\Delta T_{m^2}(x,y)$, the
corresponding formula for $\Delta P(x,y)$ is obtained by calculating 
the partial derivatives and using the commutation rules of the Dirac 
matrices in (\ref{6}), which will be a (lengthy but) straightforward
computation.

We conclude that the main problem is to calculate the solution (\ref{7}) of 
the Klein-Gordon equation (\ref{8}).
The simplest method is to analyze the partial 
differential equation (\ref{8}). This hyperbolic equation is closely related
to the wave equation, and the behavior near the light cone can
be studied like the wave propagation of singularities (this method is 
sometimes called ``integration along characteristics''; see e.g. \cite{FL}).
In order to give an idea of the technique, we look at the simplified equation
\begin{equation}
        (-\OBox - m^2) \:f(x) \;=\; g(x)
        \label{aa}
\end{equation}
and choose light-cone coordinates $(u=\frac{1}{2}(t+r),\: v=\frac{1}{2}(t-r),\:
\vartheta, \varphi)$ around the origin ($r, \vartheta, \varphi$ are 
polar coordinates in $\R^3$). Then the $\OBox$-operator takes the form
\[ \OBox \;=\; \partial_u \partial_v \:-\:\frac{1}{r} \:(\partial_u - 
\partial_v) \:-\: \frac{1}{r^2} \: 
\Delta_{S^2} \;\;\; , \]
where $\Delta_{S^2}=\partial_{\vartheta}^2 + \cot \vartheta \:
\partial_{\vartheta} + \sin^{-2} \vartheta \: \partial_{\varphi}^2$ is the
spherical Laplace operator. The important point is that the
$\OBox$-operator contains only first derivatives in both $u$ and $v$.
This allows us to express the 
normal derivative of $f$ on the light cone as a line integral over 
$f$ and its tangential derivatives. Thus,
we rewrite (\ref{aa}) on the upper light cone $u=t=r$, $v=0$, in 
the form
\[ \partial_u \left( u \: \partial_v f(u,0,\vartheta, \varphi) 
\right) \;=\; \left( \partial_u - u \:m^2 +
\frac{1}{u} \: \Delta_{S^2} \right) f(u,0,\vartheta,\varphi) \:-\:
u \:g(u,0,\vartheta,\varphi) \;\;\; . \]
This equation can be integrated along the light cone as
\begin{eqnarray}
\lefteqn{ u_1 \:\partial_v f(u_1, 0, \vartheta, \varphi) \;=\;
        \int_{0}^{u_1} \partial_u \left( u \:\partial_v f(u, 0, \vartheta, \varphi)
        \right) \; du } \nonumber \\
         & = & \int_{0}^{u_1} \left( (\partial_u - u \:m^2 +
         \frac{1}{u} \: \Delta_{S^2} )
         f(u, 0, \vartheta, \varphi) \:-\:
         u \:g(u, 0, \vartheta, \varphi) \right)\: du \nonumber \\
         & = & f(u_1,0,\vartheta,\varphi) \:-\: f(0,0,\vartheta,\varphi) 
         \nonumber \\
         &&+\int_{0}^{u_1} \left( (-u \:m^2+\frac{1}{u} \: \Delta_{S^2}) 
         f(u, 0, \vartheta, \varphi)\:-\:
         u \:g(u, 0, \vartheta, \varphi) \right) \: du \;\;\; .
         \label{0a}
\end{eqnarray}
By iterating this method, it is possible to calculate the higher 
derivatives in a similar way.
We conclude that knowing $f$ on the light cone
determines all its derivatives on the light cone.
This makes it possible to perform the light-cone expansion.
We remark that complications arise when $f$ has singularities on the light 
cone. The main disadvantage of this method is that the special form of the 
solution (\ref{7}) does not enter. This means, in our example, that 
additional input is needed to completely determine $f$ on the light 
cone.

Because of these problems, it is preferable to use a different method
and to directly evaluate the integral (\ref{7}). One substitutes
explicit formulas for the distributions $S$ and $T$ in position space 
and studies the asymptotic behavior of the integral for $(y-x)^2
\rightarrow 0$.
This method is presented in detail in \cite{F2}.
Because it is carried out purely in position space, it gives a
good intuition for the behavior of $\Delta P$ near the light cone.
Unfortunately, this method is rather lengthy. Furthermore, the calculation of 
the operator products in (\ref{7}) and of the partial derivatives in 
(\ref{6}) lead to subtle analytical difficulties.

In this paper, we use a combination of calculations in 
position and in momentum space, which gives a shorter and more
systematic approach. It has the disadvantage that working with 
infinite sums in momentum space is more abstract than studying the 
behavior of distributions in position space. Therefore the 
reader may find it instructive to compare the technique of this paper 
with the calculations in \cite{F2}.

\section{The Formal Light-Cone Expansion of $\Delta T_{m^2}$}
\setcounter{equation}{0}
In this section, we will perform the light-cone expansion for $\Delta 
T_{m^2}(x,y)$ on a formal level. 
The analytic justification for the expansion is postponed to the next 
section. We assume that $m \neq 0$ and set $a=m^2$.

Since we want to derive formulas in position space, it is useful first to 
consider explicitly how $T_{m^2}$ looks like.
Calculating the Fourier transform of the lower mass shell (\ref{8a}) 
yields an expression containing the Bessel functions $J_1$, $Y_1$,
and $K_1$. The most convenient form for our purpose is to work with the power
series for these Bessel functions, which gives
\begin{eqnarray}
        T_a(x,y) & = & -\frac{1}{8 \pi^3} \:\lim_{0<\varepsilon \rightarrow 0}
        \: \frac{1}{\xi^2-i \varepsilon \xi^0} \nonumber \\
        &&+\: \frac{a}{32 \pi^3} \:\lim_{0<\varepsilon \rightarrow 0}
        \left( \log(a \xi^2-i \varepsilon \xi^0) + i \pi + c \right)
        \sum_{l=0}^\infty \frac{(-1)^l}{l! \: (l+1)!} \: \frac{(a 
        \xi^2)^l}{4^l} \nonumber \\
         &  & -\frac{a}{32 \pi^3} \sum_{l=0}^\infty \frac{(-1)^l}{l! \: 
         (l+1)!} \: \frac{(a \xi^2)^l}{4^l} \: (\Phi(l+1) + \Phi(l))
        \label{10}
\end{eqnarray}
with $\xi=(y-x)$, $c=2C-\log 2$ with Euler's constant $C$, and the function
\[ \Phi(0)=0 \;\;\;\;,\spc \Phi(n)=\sum_{k=1}^n \frac{1}{k}
\;\;\;\;{\mbox{for}} \;\;\; n \geq 1 \;\;\; . \]
The logarithm is understood in 
the complex plane, which is cut along the positive real axis
(so that $\lim_{0<\varepsilon \rightarrow 0} \log(x+i \varepsilon) = 
\log |x|$ is real for $x>0$).
It can be verified explicitly that $T_a$ is a solution of the 
Klein-Gordon equation $(-\OBox_x-a) \:T_a(x,y)=0$. Furthermore, one can
calculate the Fourier transform $T_a(p)$ with contour integrals.
For $p^0>0$, the $\xi^0$-integral can be closed in the lower complex plane,
which gives zero.
In this way, one immediately verifies that $T_a$ is formed only of 
negative-energy states.

This formula for $T_a$ looks quite complicated, and we do not need the
details in this section. It suffices 
to observe that $T_a$ has singularities on the light cone of the form
of a pole and a $\delta$-distribution,
\[ \lim_{0<\varepsilon \rightarrow 0} \frac{1}{\xi^2-i\varepsilon 
\xi^0} \;=\; \frac{\mbox{PP}}{\xi^2} \:+\: i \pi \delta (\xi^2) \: 
\varepsilon (\xi^0) \;\;\; , \]
where PP denotes the principal value. Furthermore, there are 
logarithmic and $\Theta$-like contributions, since
\[ \lim_{0<\varepsilon \rightarrow 0} \log(a \xi^2 - i \varepsilon 
\xi^0) \:+\: i \pi \;=\; \log(|a \xi^2|) \:+\: i \pi \:\Theta(\xi^2) 
\: \epsilon(\xi^0) \;\;\; , \]
where $\epsilon$ is the step function $\epsilon(x)=1$ for $x \geq 0$ 
and $\epsilon(x)=-1$ otherwise. The important point for the following 
is the qualitative observation that the contributions of higher 
order in $a$ contain more factors $\xi^2$ and are thus of higher order 
on the light cone. This yields the possibility of
performing the light-cone expansion by expressing
$\Delta T_a$ in terms of the $a$-derivatives of $T_a$.
In the following lemma, we combine this idea with the fact that line 
integrals over the potential should occur according to (\ref{0a}).
The lemma gives an explicit solution of the
inhomogeneous Klein-Gordon equation (\ref{8}) and is the key for the 
light-cone expansion. We use the notation
\[ T_a^{(n)} \;=\; \left(\frac{d}{da}\right)^n T_a \;\;\; . \]
\begin{Lemma}
The formal series
\begin{equation}
        A(x,y) \;=\; -\sum_{n=0}^\infty \frac{1}{n!} \:
        \int_0^1 (\alpha-\alpha^2)^n \: (\OBox^n V)_{|\alpha y+(1-\alpha)x } 
        \: d\alpha \; T_a^{(n+1)}(x,y)
        \label{13}
\end{equation}
satisfies the equation
\begin{equation}
        (-\OBox_x-a) \: A(x,y) \:=\: -V(x) \: T_a(x,y) \;\;\; .
        \label{14}
\end{equation}
\end{Lemma}
{\Proof}
In momentum space, $T_a$ has the form
\[ T_a(p) \;=\; \delta(p^2-a) \: \Theta(-p^0) \;\;\; . \]
Since $a>0$, the mass shell does not intersect the hyperplane $p^0=0$, 
and we can thus calculate the distributional derivative to
\begin{eqnarray*}
        \frac{\partial}{\partial p^j} \left( \delta^{(n)}(p^2-a) \: 
        \Theta(-p^0) \right)
        & = & 2 p_j \: \delta^{(n+1)}(p^2-a) \: \Theta(-p^0) 
        \:-\: \delta^{(n)}(p^2-a) \: \delta^0_j \: \delta(p^0)  \\
         & = & 2 p_j \:\delta^{(n+1)}(p^2-a) \:\Theta(-p^0) \;\;\; .
\end{eqnarray*}
Hence, for the calculation of derivatives we can view $T_a$ as a function 
of $(p^2-a)$; i.e.
\begin{equation}
        \frac{\partial}{\partial p^j} T_a^{(n)}(p) \;=\; -2 p_j \: T_a^{(n+1)}(p)
        \;\;\; . \label{16}
\end{equation}
This relation can also be used to calculate the derivatives of 
$T_a^{(n)}$ in position space,
\begin{eqnarray}
        \frac{\partial}{\partial x^j} T_a^{(n)}(x,y) & = & \int 
        \frac{d^4p}{(2 \pi)^4} \; T_a^{(n)}(p) \; (-i p_j) \: e^{-ip(x-y)}
        \nonumber \\
         & = & \frac{i}{2} \: \int \frac{d^4p}{(2 \pi)^4} \:
         \frac{\partial}{\partial p^j} T_a^{(n-1)}(p) \; e^{-ip(x-y)} 
         \nonumber \\
         & = & -\frac{i}{2} \: \int \frac{d^4p}{(2 \pi)^4} \;
         T_a^{(n-1)}(p) \; \frac{\partial}{\partial p^j} e^{-ip(x-y)} 
         \nonumber \\
         &=& \frac{1}{2} \:(y-x)_j \: T_a^{(n-1)}(x,y) \;\;\; .
         \label{11}
\end{eqnarray}
Using that $T_a$ is a solution of the Klein-Gordon equation, we also have
\[      0 \;=\; \left(\frac{d}{da}\right)^{n} (p^2-a) \: T_a(p)
        \;=\; (p^2-a) \: T^{(n)}_a(p) \:-\: n \: T_a^{(n-1)}(p) \]
and thus
\begin{equation}
        (-\OBox_x - a) \: T^{(n)}_a(x,y) \;=\; n \: T^{(n-1)}_a(x,y) \;\;\; .
        \label{12}
\end{equation}
With the help of (\ref{11}) and (\ref{12}), we can calculate the 
derivatives of the individual summands in (\ref{13}) as follows:
\begin{eqnarray*}
\lefteqn{ (-\OBox_x-a) \int_0^1 (\alpha-\alpha^2)^n \:
(\OBox^n V)_{|\alpha y+(1-\alpha)x } \:d\alpha \; T_a^{(n+1)}(x,y) } \\
&=& -\int_0^1 (1-\alpha)^2 \: (\alpha-\alpha^2)^n \:
(\OBox^{n+1} V)_{|\alpha y+(1-\alpha)x } \:d\alpha \; T_a^{(n+1)}(x,y) \\
&&-\int_0^1 (1-\alpha) \:(\alpha-\alpha^2)^n \:
(\partial_j \OBox^n V)_{|\alpha y+(1-\alpha)x } \:d\alpha \;
(y-x)^j \:T_a^{(n)}(x,y) \\
&&+(n+1) \int_0^1 (\alpha-\alpha^2)^n \:
(\OBox^n V)_{|\alpha y+(1-\alpha)x } \:d\alpha \; T_a^{(n)}(x,y) \;\;\; .
\end{eqnarray*}
In the second summand, we rewrite the partial derivative as a
derivative with respect to $\alpha$,
\[ (y-x)^j \:\partial_j \OBox^n V_{|\alpha y+(1-\alpha)x} \;=\;
\frac{d}{d\alpha} \OBox^n V_{|\alpha y+(1-\alpha)x} \;\;\; , \]
and integrate by parts, which gives
\begin{eqnarray*}
\lefteqn{ (-\OBox_x-a) \int_0^1 (\alpha-\alpha^2)^n \:
(\OBox^n V)_{|\alpha y+(1-\alpha)x } \:d\alpha \; T_a^{(n+1)}(x,y) } \\
&=& \delta_{n,0} \: V(x) \; T_a(x,y) \\
&&+n \int_0^1 (1-\alpha)^2 \: (\alpha-\alpha^2)^{n-1} \:
(\OBox^{n} V)_{|\alpha y+(1-\alpha)x } \:d\alpha \; T_a^{(n)}(x,y) \\
&&-\int_0^1 (1-\alpha)^2 \: (\alpha-\alpha^2)^{n} \:
(\OBox^{n+1} V)_{|\alpha y+(1-\alpha)x } \:d\alpha \; T_a^{(n+1)}(x,y) \;\;\; .
\end{eqnarray*}
After dividing by $n!$ and summation over $n$, the last two summands 
are telescopic and vanish. This yields (\ref{14}).
\QED
It would be nice if the solution of the
inhomogeneous Klein-Gordon equation constructed 
in the previous lemma coincided with $\Delta T_a(x,y)$.
This is really the case, although it is not obvious. In the remainder of 
this section, we will prove it. The technique is to
expand (\ref{7}) in momentum space and to show that the resulting 
expression is the Fourier transform of (\ref{13}).

Since (\ref{7}) is linear in $V$, we can assume that $V$ has the form 
of a plane wave,
\begin{equation}
        V(x) \;=\;  e^{-iqx} \;\;\; .
        \label{e}
\end{equation}
Transforming $\Delta T_a(x,y)$ to momentum space gives the formula
\begin{eqnarray}
\lefteqn{ \int d^4x \int d^4y \; \Delta T_a(x,y) \: e^{ip_2y - i 
p_1x} } \nonumber \\
&=&-\left( \frac{{\mbox{PP}}}{p_2^2-a} \: T_a(p_1) \:+\:
         T_a(p_2) \: \frac{{\mbox{PP}}}{p_1^2-a} \right) \: 
         \delta^4(q-p_2+p_1) \;\;\; , \label{37n}
\end{eqnarray}
where $p_1, p_2$ are the in- and outgoing momenta, and where PP denotes the
principal value,
\[ \frac{\mbox{PP}}{p^2-a} \;=\; \frac{1}{2} \:\lim_{0<\varepsilon 
\rightarrow 0} \sum_{\pm} \frac{1}{p^2-a \pm i \varepsilon p^0} 
\;\;\; . \]
The factor $\delta^4(q-p_2+p_1)$ in (\ref{37n}) describes the conservation of 
energy-momentum. In order to simplify the notation, we leave out this 
$\delta^4$-factor and view $\Delta T_a$ as a function of only one free variable 
$p=(p_1+p_2)/2$,
\begin{equation}
        \Delta T_a(p+\frac{q}{2}, p-\frac{q}{2}) \;:=\; -\frac{{\mbox{PP}}}
        {(p+\frac{q}{2})^2-a} \: T_a(p-\frac{q}{2}) \:-\:
         T_a(p+\frac{q}{2}) \: \frac{{\mbox{PP}}}{(p-\frac{q}{2})^2-a}
         \;\;\; .       \label{15}
\end{equation}
The transformation to position space is then given by a single integral,
\begin{equation}
         \Delta T_a(x,y) \;=\; \int \frac{d^4p}{(2 \pi)^4} \: \Delta
         T_a(p+\frac{q}{2}, p-\frac{q}{2}) \; e^{-ip(x-y)} \:
         e^{-i \frac{q}{2} \:(x+y)} \;\;\; ,
        \label{d}
\end{equation}
the inverse transformation to momentum space takes the form
\begin{equation}
\Delta T_a(p+\frac{q}{2}, p-\frac{q}{2}) \;=\; \int d^4y \; \Delta 
T_a(x,y) \; e^{ip(x-y)} \:e^{i \frac{q}{2} \:(x+y)} \;\;\; .
        \label{39a}
\end{equation}

The first step for the light-cone expansion in momentum space is 
to rewrite (\ref{15}) in the form
\[ \Delta T_a(p+\frac{q}{2}, p-\frac{q}{2}) \;=\; 
-\frac{\mbox{PP}}{[(p-\frac{q}{2})^2-a] + 2pq} \: T_a(p-\frac{q}{2}) 
\:-\: T_a(p+\frac{q}{2}) \: \frac{\mbox{PP}}{[(p+\frac{q}{2})^2-a] - 
2pq} \]
and to use that the expressions within brackets $[\cdots]$ vanish as the 
arguments of the $\delta$-distributions in $T_a(p \pm \frac{q}{2})$,
\[ \Delta T_a(p+\frac{q}{2}, p-\frac{q}{2}) \;=\;
\frac{\mbox{PP}}{2pq} \: \left( T_a(p+\frac{q}{2}) - T_a(p-\frac{q}{2})
\right) \;\;\; . \]
Now we expand $T_a$ in a Taylor series in $q$,
\begin{equation}
\Delta T_a(p+\frac{q}{2}, p-\frac{q}{2}) \;=\;
\frac{\mbox{PP}}{2pq} \; 2 \sum_{k=0}^\infty \frac{1}{(2k+1)!} \:
\frac{\partial^{2k+1}}{\partial p^{i_1} \cdots \partial p^{i_{2k+1}}}
T_a(p) \; \frac{q^{i_1}}{2} \cdots \frac{q^{i_{2k+1}}}{2} \;\;\; .
        \label{b}
\end{equation}
We want to rewrite the $p$-derivatives as derivatives with 
respect to $a$. This can be done by iteratively carrying out the
$p$-derivatives with the differentiation rule (\ref{16}). One must keep in mind
that the $p$-derivatives act either on $T_a$ or on the factors $p_j$ that
were previously generated by (\ref{16}), e.g.,
\[ \frac{\partial^2}{\partial p^j \partial p^k} T_a(p) \:\frac{q^j}{2} 
\frac{q^k}{2} \;=\; -\frac{\partial}{\partial p^j} \left( (pq) 
\:T_a^{(1)}(p) \right) \frac{q^j}{2} \;=\; (pq)^2 \:T_a^{(2)}(p) 
\:-\:\frac{q^2}{2} \:T_a^{(1)}(p) \;\;\; . \]
In this way, carrying out the $p$-derivatives in (\ref{b}) gives a sum of many
terms. The combinatorics may be described as follows: Each factor $p_j$ that
is generated by (\ref{16}) and differentiated thereafter gives a pairing
between 
two of the derivatives $\partial_{i_1},\ldots, \partial_{i_{2k+1}}$;
namely, between the derivative by which it was created and the 
derivative by which it was subsequently annihilated.
The individual expressions obtained after carrying out all the derivatives
correspond to the possible configurations of the pairings among the
$\partial_{i_1} \cdots \partial_{i_{2k+1}}$. They depend only on the number
of pairs and not on their specific 
configuration. More precisely, every pair increases the degree of the 
derivative of $T_a^{(.)}$ by one and gives a factor $q^2/2$, whereas 
the unpaired derivatives also increase the degree of $T_a^{(.)}$ and 
give a factor $pq$.

It remains to count how many configurations of such $n$ pairs exist.
We use the notation $\left[ \!\!\begin{array}{c} m \\ n \end{array} 
\!\! \right]$ for the number of possibilities to choose $n$ pairs from 
a set of $m \geq 2n$ points.
The combinatorics becomes clearer if one first selects $2n$ out of 
the $m$ points and then counts the number of possible pairings among 
these $2n$ points to $(2n-1)!!$. This explains the formula
\begin{equation}
        \left[\!\! \begin{array}{c} m \\ n \end{array} 
        \!\!\right] \;=\;
        \left(\!\! \begin{array}{c} m \\ 2n \end{array} 
        \!\!\right) \: (2n-1)!! \;=\; \frac{m!}{(m-2n)!} \: 
        \frac{1}{2^n \: n!} \;\;\; .
        \label{25z}
\end{equation}
We conclude that
\begin{eqnarray*}
\lefteqn{ \frac{\partial^{2k+1}}{\partial p^{i_1} \cdots \partial p^{i_{2k+1}}}
        T_a(p) \; \frac{q^{i_1}}{2} \cdots \frac{q^{i_{2k+1}}}{2} } \\
        &=& \sum_{n=0}^k (-1)^{1+n} \:\frac{(2k+1)!}{(2k+1-2n)!} \: \frac{1}{2^n \: n!} 
        \; T_a^{(2k+1-n)}(p) \: \left(\frac{q^2}{2}\right)^n \:
        (pq)^{2k+1-2n} \;\;\; .
\end{eqnarray*}
After substituting into (\ref{b}) and reordering the sums, we obtain
\begin{equation}
        \Delta T_a(p+\frac{q}{2}, p-\frac{q}{2}) \;=\; -\frac{\mbox{PP}}{pq}
        \sum_{n=0}^\infty \frac{(-1)^n}{n!} \left(\frac{q^2}{4}\right)^n
        \sum_{k=0}^\infty \frac{(pq)^{2k+1}}{(2k+1)!} \: T_a^{(2k+1+n)}(p) \;\;\; .
        \label{25a}
\end{equation}
Finally, we pull one factor $pq$ out of the sum, which cancels the 
principal value,
\begin{equation}
        \Delta T_a(p+\frac{q}{2}, p-\frac{q}{2}) \;=\;
        -\sum_{n=0}^\infty \frac{(-1)^n}{n!} \left(\frac{q^2}{4}\right)^n
        \sum_{k=0}^\infty \frac{(pq)^{2k}}{(2k+1)!} \: T_a^{(2k+1+n)}(p) 
        \;\;\; .
        \label{c}
\end{equation}
This is the formula for the light-cone expansion in momentum space.

It remains to show that the Fourier transform (\ref{d}) of (\ref{c}) 
coincides with (\ref{13}). In order to see a first similarity between 
these formulas, we substitute the plane-wave ansatz (\ref{e}) into 
(\ref{13}):
\begin{equation}
        A(x,y) \;=\; -\sum_{n=0}^\infty \frac{(-q^2)^n}{n!}
        \int_{-\frac{1}{2}}^{\frac{1}{2}} (\frac{1}{4} - \tau^2)^n \:
        e^{-i\tau q (y-x)} \:d\tau \; T_{a}^{(n+1)}(x,y) \; 
        e^{-i\frac{q}{2} \: (x+y)} \;\;\; .
        \label{f}
\end{equation}
The last exponential factor also occurs in (\ref{d}).
Furthermore, it is encouraging that both (\ref{c}) and (\ref{f}) 
contain a power series in $q^2$. The main difference between the 
formulas is related to the factor $\exp (-i\tau q(y-x))$ in (\ref{f}):
Expanding this exponential yields a power series in $q(y-x)$. In momentum 
space, this corresponds to a power series in $q^j \partial_{p^j}$ 
(because differentiating the exponential in (\ref{d}) with respect to 
$p$ gives a factor $i(y-x)$). The expansion (\ref{c}), however, 
contains a power series in $pq$, not in $q^j \partial_{p^j}$.
The following lemma allows us to transform these expansions into each 
other.
\begin{Lemma}
\label{lemma_convert}
For all $r \geq 0$,
\begin{eqnarray}
        \left(\frac{q^j}{2} \frac{\partial}{\partial p^j}\right)^{2k}
        T_a^{(r)}(p) & = & \sum_{l=0}^k (-1)^l \:\left[\!\! \begin{array}{c} 2k \\ l
        \end{array} \!\!\right] \:\left( \frac{q^2}{2} \right)^l \:
        (pq)^{2k-2l} \:T_a^{(r+2k-l)}(p)
        \label{20}  \\
        (pq)^{2k} \: T_a^{(r+2k)}(p) & = & \sum_{l=0}^k
        \left[\!\! \begin{array}{c} 2k \\ l
        \end{array} \!\!\right] \:\left( \frac{q^2}{2} \right)^l \:
        \left(\frac{q^j}{2} \frac{\partial}{\partial p^j}\right)^{2k-2l}
        T_a^{(r+l)}(p) \label{21} \;\;\; .
\end{eqnarray}
\end{Lemma}
{\Proof}
On the left side of (\ref{20}), we calculate the derivatives inductively 
using the differentiation rule (\ref{16}). The derivatives either act on
$T_a^{(.)}$,
which increases the order of the derivative of $T_a^{(.)}$ and generates a 
factor $pq$, or they act on previously generated factors $pq$, which reduces
the number of factors $pq$ by one and produces a factor $q^2/2$.
This can also be written in the inductive form
\[ \left(\frac{q^j}{2} \frac{\partial}{\partial p^j}\right)^l
T_a^{(s)}(p) \;=\; -pq \: \left(\frac{q^j}{2} \frac{\partial}
{\partial p^j}\right)^{l-1} T_a^{(s+1)}(p) \:-\:
(l-1) \:\frac{q^2}{2} \:\left(\frac{q^j}{2} \frac{\partial}
{\partial p^j}\right)^{l-2} T_a^{(s+1)}(p) \;\;\; . \]
The combinatorics is described by counting the 
number of possibilities in forming $l$ pairs among the $2k$ 
derivatives.

Equation (\ref{21}) follows in the same way from the relation
\[ (pq)^l \: T_a^{(s+l)}(p)
\;=\; -\left(\frac{q^j}{2} \frac{\partial}{\partial p^j}\right)
         (pq)^{l-1} \:T_a^{(s+l-1)}(p) \:-\: (l-1)
         \:\frac{q^2}{2} \: (pq)^{l-2} \: T^{(s+l-1)}(p) 
         \;\;\; . \;\;\; \FBox\]
After these preparations, we can prove the main result of this section:
\begin{Thm}{\bf{(formal light-cone expansion of $\Delta T_{m^2}$)}}
\label{thm1}
For $m \neq 0$, the distribution $\Delta T_{m^2}(x,y)$, (\ref{7}), has 
a representation as the formal series
\begin{equation}
\Delta T_{m^2}(x,y) \;=\; -\sum_{n=0}^\infty \frac{1}{n!} \:
        \int_0^1 (\alpha-\alpha^2)^n \: (\OBox^n V)_{|\alpha y+(1-\alpha)x } 
        \: d\alpha \; T_{m^2}^{(n+1)}(x,y) \;\;\; .
        \label{22}
\end{equation}
\end{Thm}
{\Proof}
We expand the factor $(\frac{1}{4}-\tau^2)^n$ and the exponential
$\exp(-i \tau q(y-x))$ in (\ref{f}), and so have
\begin{eqnarray}
A(x,y) &=& -\sum_{n=0}^\infty \frac{(-q^2)^n}{n!}
        \int_{-\frac{1}{2}}^{\frac{1}{2}}
        \sum_{l=0}^n \left(\!\! \begin{array}{c} n \\ l \end{array} \!\!\right)
        (-\tau^2)^l \: \left(\frac{1}{4}\right)^{n-l} \nonumber \\
&&\hspace*{4cm} \times \; e^{-i\tau q (y-x)} \:d\tau \; T_{a}^{(n+1)}(x,y) \; 
        e^{-i\frac{q}{2} \: (x+y)} \label{31p} \\
&=& -\sum_{n=0}^\infty \frac{(-q^2)^n}{n!}
        \sum_{k=0}^\infty \frac{(-iq(y-x))^{k}}{k!}
        \sum_{l=0}^n (-1)^l \left(\!\! \begin{array}{c} n \\ l \end{array} \!\!\right)
        \left(\frac{1}{4}\right)^{n-l} \nonumber \\
&&\hspace*{4cm} \times \; \int_{-\frac{1}{2}}^{\frac{1}{2}}
        \tau^{k+2l} \:d\tau \; T_{a}^{(n+1)}(x,y) \; 
        e^{-i\frac{q}{2} \: (x+y)} \;\;\; . \nonumber
\end{eqnarray}
Next we carry out the $\tau$-integration. This gives a 
contribution only for $k$ even,
\begin{eqnarray*}
&=& -\sum_{n=0}^\infty \frac{(-q^2)^n}{n!}
\sum_{k=0}^\infty \frac{(-iq(y-x))^{2k}}{(2k)!}
        \sum_{l=0}^n (-1)^l \left(\!\! \begin{array}{c} n \\ l \end{array} \!\!\right)
        \left(\frac{1}{4}\right)^{n-l} \:
        \frac{T_{a}^{(n+1)}(x,y)}{(2k+2l+1) \:4^{k+l}} \; e^{-i\frac{q}{2} \: 
        (x+y)} \\
&=& -\sum_{n=0}^\infty
                \frac{(-1)^n}{n!} \left(\frac{q^2}{4} \right)^n \sum_{k=0}^\infty
                \frac{(-iq(y-x))^{2k}}{4^k \:(2k)!} \sum_{l=0}^n \frac{(-1)^l}{2k+2l+1} \:
                \left(\!\! \begin{array}{c} n \\ l \end{array} \!\!\right)
                \; T_a^{(n+1)}(x,y) \:e^{-i \frac{q}{2} (x+y)} \;\;\; .
\end{eqnarray*}
Now we transform to momentum space by substituting into (\ref{39a}). The factors 
$-iq(y-x)/2$ can be rewritten as derivatives $\frac{q^j}{2} 
\partial_{p^j}$ acting on the plane wave $e^{ip(x-y)}$. Integrating 
these $p$-derivatives by parts gives
\[ A(p+\frac{q}{2}, p-\frac{q}{2}) \:=\: -\sum_{n=0}^\infty
        \frac{(-1)^n}{n!} \left(\frac{q^2}{4} \right)^n \sum_{k=0}^\infty
        \frac{1}{(2k)!} \sum_{l=0}^n \frac{(-1)^l}{2k+2l+1} \:
        \left(\!\! \begin{array}{c} n \\ l \end{array} \!\!\right) \:
        \left(\frac{q^j}{2} \frac{\partial}{\partial p^j}\right)^{2k}
        T_a^{(n+1)}(p) \;\; . \]
We shift the indices $n$ and $k$ according to $n-l \rightarrow n$ and 
$k+l \rightarrow k$. This changes the range of the $l$-summation to 
$l=0,\ldots,k$. We thus obtain
\begin{eqnarray}
\lefteqn{ \Delta T_a(p+\frac{q}{2}, p-\frac{q}{2}) } \nonumber \\
&=& -\sum_{n=0}^\infty
        \frac{(-1)^n}{n!} \left( \frac{q^2}{4} \right)^n \sum_{k=0}^\infty
        \frac{1}{2k+1} \sum_{l=0}^k
        \frac{1}{(2k-2l)! \; l!} \:\left( \frac{q^2}{4} \right)^l \:
        \left(\frac{q^j}{2} \frac{\partial}{\partial p^j}\right)^{2k-2l}
        T_a^{(n+1+l)}(p) \nonumber \\
&=& -\sum_{n=0}^\infty
        \frac{(-1)^n}{n!} \left( \frac{q^2}{4} \right)^n \sum_{k=0}^\infty
        \frac{1}{(2k+1)!} \sum_{l=0}^k
        \left[\!\! \begin{array}{c} 2k \\ l
        \end{array} \!\!\right] \:\left( \frac{q^2}{2} \right)^l \:
        \left(\frac{q^j}{2} \frac{\partial}{\partial p^j}\right)^{2k-2l}
        T_a^{(n+1+l)}(p) \;. \;\;\;\;\;\;\;
        \label{3e}
\end{eqnarray}
In this way, we have transformed the line integrals in (\ref{f}) to 
momentum space. We remark that we did not use the special form of 
$T_a$ in the calculation from (\ref{31p}) to (\ref{3e}); for the computation
so far, we could replace $T_a$ by any other function.

In the last step, we apply Lemma \ref{lemma_convert}: substituting (\ref{21})
into the light-cone expansion (\ref{c}) for $\Delta T_a$ also yields the 
expression (\ref{3e}). Thus $A$ coincides 
with $\Delta T_a$, which concludes the proof.
\QED

\section{Resummation of the Noncausal Contribution}
\setcounter{equation}{0}
In this section, we will put the previous formal calculations on a 
rigorous basis. The interesting part is to recover the noncausal 
structure of $\Delta T_{m^2}(x,y)$ by resumming the formal light-cone 
expansion. We begin with specifying the conditions on the potential $V$
in (\ref{7}).
\begin{Lemma}
Let $V \in L^1(\R^4)$ be a potential which decays so fast at infinity 
that the functions $x^i \: V(x)$ are also $L^1$. Then $\Delta 
T_{m^2}(x,y)$ of (\ref{7}) is a well-defined tempered distribution on
$\R^4 \times \R^4$.
\end{Lemma}
{\Proof}
It is easier to proceed in momentum space and to show that
\begin{equation}
        \Delta T_{m^2}(p_2, p_1) \;=\; -S_{m^2}(p_2) \:\tilde{V}(p_2-p_1) 
        \:T_{m^2}(p_1) \:-\: T_{m^2}(p_2) \:\tilde{V}(p_2-p_1) 
        \:S_{m^2}(p_1)
        \label{40}
\end{equation}
is a well-defined tempered distribution, where $\tilde{V}$ is the 
Fourier transform of $V$. The assumption then follows by Fourier 
transformation.

In momentum space, the conditions on the potential give
$\tilde{V} \in C^1(\R^4) \cap L^\infty(\R^4)$. We choose two test 
functions $f,g \in S(\R^4)$. Then the function $\tilde{V}(p_2-.) \: 
f(.)$ is $C^1$ and has rapid decay at infinity. Thus the 
integral over the lower mass shell
\[ I(p_2) \;:=\; \int \frac{d^4 p_1}{(2 \pi)^4} \;
\tilde{V}(p_2-p_1) \: T_{m^2}(p_1) \: f(p_1) \]
is finite and depends differentiably on $p_2$. Consequently, the 
product $g I$ is in $C^1$ and has rapid decay, and we can 
calculate the principal value by
\[ \frac{1}{2} \:\lim_{0<\varepsilon \rightarrow 0} \sum_\pm \int 
\frac{d^4p_2}{(2 \pi)^4} \: \frac{1}{p_2^2-m^2\pm i \varepsilon p_2^0} \:
g(p_2) \: I(p_2) \;\;\; , \]
which gives a finite number. This shows that the first summand
\[ S_{m^2}(p_2) \:\tilde{V}(p_2-p_1) \:T_{m^2}(p_1) \]
in (\ref{40}) is a well-defined linear functional on $S(\R^4) \times S(\R^4)$.
This functional is bounded in the Schwartz norms $\|.\|_{0,0}$,
$\|.\|_{4,0}$, $\|.\|_{0,1}$, $\|.\|_{4,1}$
of $f, g$, which gives continuity.

For the second summand in (\ref{40}), one can argue in the same way after 
exchanging $p_1$ and $p_2$.
\QED
For the light-cone expansion, we clearly need a smooth potential. 
Therefore, we will assume in the following that $V \in C^\infty \cap L^1$ 
and $x^i \:V(x) \in L^1$.

We come to the mathematical analysis of the light-cone expansion. We again
assume that $m \neq 0$; the case $m=0$ will be obtained at the end of this 
section in the limit $m \rightarrow 0$. In the first step, we 
disregard the convergence of the infinite sums and check that all the 
performed operations make sense and that all expressions are well-defined.
We start with the end formula (\ref{22}) of the light-cone expansion.
The line integrals over the potentials are $C^\infty$-functions in 
$x, y$. The factors $T^{(n)}_{m^2}(x,y)$ are tempered distributions, 
as one sees after differentiating the explicit formula (\ref{10})
with respect to $a$. Thus (\ref{22}) makes mathematical sense.
The calculations leading to this result are not problematic except for
the handling of the principal value following (\ref{15}).
The easiest method for studying this more
rigorously is to regularize the principal value 
in (\ref{15}) with the replacement
\[ \frac{\mbox{PP}}{x} \;\longrightarrow\; \frac{1}{2} \sum_{\pm}
\frac{1}{x\pm i \varepsilon} \;=\; \frac{x}{x^2+\varepsilon^2} 
\;\;\; . \]
Then all the subsequent transformations are well-defined, and the 
critical operation is the cancellation of the principal 
value against one factor $pq$ before (\ref{c}).
In order to justify this operation, we use in (\ref{25a}) the exact 
formula
\begin{eqnarray*}
\lefteqn{ \frac{pq}{(pq)^2+\varepsilon^2} \: (pq)^{2k+1} \;=\;
        \frac{((pq)^2+\varepsilon^2) - \varepsilon^2}{(pq)^2+\varepsilon^2}
        \: (pq)^{2k} \;=\; \left(1 - \frac{\varepsilon^2}{(pq)^2+\varepsilon^2}
        \right) (pq)^{2k} } \\
         & = & \cdots \;=\; (pq)^{2k} \:-\: \varepsilon^2 \:(pq)^{2k-2} \:+\:
         \cdots\:+\:(-1)^k \: \varepsilon^{2k} \:+\: (-1)^{k+1} \:
         \frac{\varepsilon^{2k+2}}{(pq)^2+\varepsilon^2} \;\;\; .
\end{eqnarray*}
The first summand gives (\ref{c}); the following summands $(-1)^l \:
\varepsilon^{2l} \: (pq)^{2k-2l}$ contain no principal value and 
vanish in the limit $\varepsilon \rightarrow 0$. Thus it remains to consider
the last summand for $\varepsilon \rightarrow 0$,
\begin{equation}
        \lim_{\varepsilon \rightarrow 0} 
        \frac{\varepsilon^{2k+2}}{(pq)^2+\varepsilon^2} \:\left(\frac{q^2}{4}
        \right)^n \: T_{m^2}^{(2k+1+n)}(p) \;\;\; .
        \label{21a}
\end{equation}
We use that the support of $T_{m^2}(p)$ is on the mass shell 
$p^2=m^2$ and apply the relation $\lim_{\varepsilon \rightarrow 0}
\varepsilon/(x^2+\varepsilon^2)=\pi \delta(x)$,
\begin{eqnarray*}
        \lim_{\varepsilon \rightarrow 0} 
        \frac{\varepsilon}{(pq)^2+\varepsilon^2} \: T_a^{(2k+1+n)}(p) & = & 
        \left(\frac{d}{da}\right)^{2k+1+n} \:\lim_{\varepsilon \rightarrow 0}
        \frac{\varepsilon}{(pq)^2 + \varepsilon^2} \:T_a(p)  \\
         & = & \left(\frac{d}{da}\right)^{2k+1+n} \left( \frac{\pi}{\sqrt{a}}
         \:\delta(q \:\frac{p}{|p|}) \;T_a(p) \right) \;\;\; .
\end{eqnarray*}
This expression is well-defined for $m \neq 0$. The limit (\ref{21a}) contains
an additional factor $\varepsilon^{2k+1} (q^2/4)^n$ and thus vanishes.

We conclude that the light-cone expansion is mathematically 
rigorous except for the formal character of the infinite sums. In the 
remainder of this section, we will carefully analyze the infinite sum in 
(\ref{22}). More precisely, we will do the following:
As explained in the introduction, the infinite sum in (\ref{6a}) is 
only a notation for the approximation (\ref{6b}) of the partial
sums. Following this definition, we need only show that the light-cone 
expansion is well-defined to any order on the light cone, but we need 
not study the convergence of the sum over the order on the light 
cone. According to the explicit formula (\ref{10}), each factor
$T_{m^2}^{(n+1)}$ in (\ref{22}) consists of an infinite number of terms of
different order on the light cone (we will see this in more detail in 
a moment). In order to bring (\ref{22}) into the required form (\ref{6a}), we 
must collect all contributions to a given order on the light cone and 
form their sum. This procedure is called {\em{resummation}} of the light-cone 
expansion. If the sum over all contributions to a given order on the 
light cone were finite, this resummation would be trivial; it would just
correspond to a rearrangement of the summands. It will turn out, however,
that these sums are infinite, and we must find a way to carry them out.

In order to see the basic problem in more detail, we consider the explicit 
formula (\ref{10}) for $T_a$. We start with the last summand
\begin{equation}
        T_a(x,y) \;\asymp\; -\frac{a}{32 \pi^3} \sum_{l=0}^\infty \frac{(-1)^l}
        {l! \: (l+1)!} \: \frac{(a \xi^2)^l}{4^l} \: (\Phi(l+1) + \Phi(l))
        \label{43}
\end{equation}
(the notation `$\asymp$' means that we consider only a certain 
contribution to $T_a$).
This is a power series in $a$, and we can calculate its derivatives to
\begin{equation}
        T_a^{(n)}(x,y) \;\asymp\; \frac{1}{16 \pi^3} \sum_{l=n-1}^\infty \frac{(-1)^l}
        {l! \: (l+1-n)!} \: \frac{a^{l+1-n} \: \xi^{2(l+n-1)}}{4^l} \:
        (\Phi(l+1) + \Phi(l)) \;\;,\;\;\;\; n \geq 1 \; .
        \label{44}
\end{equation}
For increasing $n$, the derivatives are of higher order on the light cone; more 
precisely, the contribution (\ref{44}) is of the order ${\cal{O}}((y-x)^{2(n-1)})$.
Thus, the contribution of (\ref{43}) to the formal light-cone expansion 
(\ref{22}) consists, to any order on the light cone, of only a finite 
number of terms. Thus the resummation is trivial. Of course, we 
could rearrange the sum by collecting all the summands 
in (\ref{43}) and (\ref{44}) of a given degree in $\xi^2$ and writing them 
together, but this is only a matter of taste and is not really needed.

For the second summand in (\ref{10}),
\begin{eqnarray}
        T_a(x,y) & \asymp & \frac{a}{16 \pi^3} \:
        \left( \log(\xi^2-i 0 \xi^0) + i \pi + c \right)
        \sum_{l=0}^\infty \frac{(-1)^l}{l! \: (l+1)!} \: \frac{(a 
        \xi^2)^l}{4^l} \label{45}  \\
         &  & +\frac{a}{16 \pi^3} \:\log a \:
        \sum_{l=0}^\infty \frac{(-1)^l}{l! \: (l+1)!} \: \frac{(a \xi^2)^l}{4^l}
        \label{46} \;\;\; ,
\end{eqnarray}
the situation is more complicated: The contribution (\ref{45}) is 
a power series in $a$ and can be discussed exactly as (\ref{43}).
The only difference (apart from the missing factors $\Phi$) is the 
prefactor $(\log(\xi^2-i 0 \xi^0) + i \pi + c)$, which has a 
logarithmic pole on the light cone. The contribution (\ref{46}), 
however, contains a factor $\log a$ and is {\em{not}}
a power series in $a$. As a consequence, the higher $a$-derivatives of (\ref{46})
are not of higher order on the light cone. For example, the contribution 
to the order ${\cal{O}}((y-x)^2)$ has the form
\begin{eqnarray*}
        T_a(x,y) & \asymp & \frac{1}{16 \pi^3} \: a \:\log a \:+\:
        {\cal{O}}((y-x)^2) \\
        T_a^{(1)}(x,y) & \asymp & \frac{1}{16 \pi^3} \:(1+\log a)
        \:+\: {\cal{O}}((y-x)^2)  \\
        T_a^{(n)}(x,y) & \asymp & \frac{(-1)^n}{16 \pi^3} \:(n-2)! \: 
        \frac{1}{a^{n-1}} \:+\: {\cal{O}}((y-x)^2) \;\;\;\;,\;\;\;\;\; n \geq 2 \;\;\; .
\end{eqnarray*}
This means that we must resum an infinite number of terms; more 
precisely,
\begin{eqnarray}
\lefteqn{ \Delta T_a^{[0]}(x,y) \;\asymp\; -\frac{1}{16 \pi^3} \int_0^1
        V_{|\alpha y + (1-\alpha) x} \: d\alpha \; (1+\log a) \nonumber } \\
        &&+\frac{1}{16 \pi^3} \sum_{n=1}^\infty \frac{1}{n} \int_0^1 
        (\alpha-\alpha^2)^n \: (\OBox^n V)_{|\alpha y + (1-\alpha) x} \: 
        d\alpha \; \frac{(-1)^n}{a^n} \:+\: \cdots \:+\: {\cal{O}}((y-x)^2) 
        \;\;\; . \;\;\; \label{47}
\end{eqnarray}
This is a serious problem. Namely, we can expect the series in (\ref{47})
to converge only if the derivatives $\OBox^n V$ do not grow too fast in the 
order $2n$ of the derivative. It turns out that analyticity of $V$ is 
necessary for convergence, which is too restrictive.

On a technical level, this convergence problem of the contributions
to $\Delta T_{m^2}$ to a given order on the light cone is a 
consequence of the factor $\log a$ in (\ref{46}); we call it the 
{\em{logarithmic mass problem}}.
Because $\Delta T_{m^2}(x,y)$ is well-defined by (\ref{6}), it is not a
problem of the perturbation expansion, but rather shows that the light-cone 
expansion was not performed properly. The deeper reason for the 
convergence problem is that we expressed $\Delta T_{m^2}(x,y)$ only 
in terms of the potential and its derivatives along the line segment
$\overline{xy}$. However, the perturbation $\Delta T_{m^2}(x,y)$ is not causal 
in this sense; it depends on $V$ in the whole Minkowski 
space (this becomes clear in (\ref{6}) from the fact that the support of 
$T_{m^2}(x,.)$ is $\R^4$). In a formal expansion, we can express 
$\Delta T_{m^2}(x,y)$ in terms of $\OBox^n V_{|\lambda y + (1-\lambda)x}$,
$0\leq \lambda \leq 1$, but we cannot expect this expansion to converge.
The simplest one-dimensional analog of this situation is the formal Taylor 
series
\[ f(x) \;=\; \sum_{n=0}^\infty \frac{1}{n!} \:f^{(n)}(0) \: x^n \]
of a smooth function. The right side cannot in general converge, 
because it is not possible to express $f(x)$, $x \neq 0$, in terms of $f^{(n)}(0)$.

The solution to the logarithmic mass problem is to reformulate the 
problematic contribution of (\ref{46}) to the light-cone expansion
(\ref{22}) as a noncausal term that is obviously finite. In some sense,
we will simply reverse our former construction of the light-cone expansion.
This is not trivial, however, because the differentiation rule (\ref{16}),
which was crucial
for rewriting the Taylor expansion (\ref{b}) as an expansion in the mass 
parameter $a$, is not valid for (\ref{46}).
In the end, we want to write the light-cone expansion in a way which 
shows that part of the behavior of $\Delta T_{m^2}(x,y)$ can be 
described with line integrals of the form (\ref{22}) whereas 
other contributions are noncausal in a specific way.

We work in momentum space. The Fourier transform of the problematic 
series (\ref{46}) is
\begin{equation}
J_a(p) \;=\; \int d^4x \;e^{ipx} \:\frac{1}{16 \pi^3} \sum_{l=0}^\infty
\frac{(-1)^l}{l! (l+1)!} \:\frac{a \xi^2}{4^l} \;=\;
\pi \sum_{l=0}^\infty \frac{a^l}{4^l \: l! \: (l+1)!} \: \OBox^l 
\delta^4(p) \;\;\; .
        \label{4a}
\end{equation}
Notice that this expression is highly singular at $p=0$; especially, 
it is not a distribution. However, it is well-defined as a distribution on 
analytic functions in $p$. This comprises all functions with compact 
support in position space, which is a sufficiently large function 
space for the following. Furthermore, we introduce the series
\begin{equation}
        L_a(p) \;=\; \pi \sum_{l=0}^\infty \frac{a^l}{4^l \: (l!)^2} \:
        \OBox^l \delta^4(p)
        \label{4b}
\end{equation}
and set
\[ J_a^{(n)} \;=\; \left(\frac{d}{da}\right)^n J_a \;\;\;,\;\;\;
        L_a^{(n)} \;=\; \left(\frac{d}{da}\right)^n L_a \;\;\; . \]
\begin{Lemma}
\label{lemma10}
The series (\ref{4a}) and (\ref{4b}) satisfy the relations
\begin{eqnarray}
        J_a(p) & = & \int_0^1 L_{\tau a}(p) \: d\tau
        \label{4c}  \\
        \frac{\partial}{\partial p^j} L_a^{(n)}(p) & = & -2 p_j \: 
        L_a^{(n+1)}(p) \;\;\; .
        \label{4d}
\end{eqnarray}
\end{Lemma}
{\Proof}
Equation (\ref{4c}) is verified by integrating the power series 
(\ref{4b}) and comparing with (\ref{4a}). The 
distribution $p_j \:\delta^4(p)$ vanishes identically. Since the derivatives of 
distributions are defined in the weak sense, it follows that
\[ 0 \;=\; \OBox^{n+1} \left( p_j \:\delta^4(p) \right) \;=\;
p_j \:\OBox^{n+1}\delta^4(p) \:+\: 2(n+1)\:\frac{\partial}{\partial 
p^j} \OBox^n \delta^4(p) \;\;\; , \]
and thus
\[ \frac{\partial}{\partial p^j} \OBox^n \delta^4(p) \;=\; 
-\frac{1}{4(n+1)} \:(\OBox^{n+1} \delta^4(p)) \; 2 p_j \;\;\; . \]
Applying this relation to every term of the series (\ref{4b}) yields that
$\partial_{p^j} L_a = -2 \:p_j L_a^{(1)}$, and (\ref{4d}) follows by
differentiating with respect to $a$.
\QED
The function $L_a$ is useful because (\ref{4d}) coincides with the 
differentiation rule (\ref{16}) for $T_a$. This implies that all the 
formulas for $T_a$, especially the manipulations of Lemma 
\ref{lemma_convert}, are also valid for $L_a$.

The following technical lemma is the key for handling the logarithmic mass 
problem.
\begin{Lemma} {\bf{(resummation of the noncausal contribution)}} If 
$V$ is the plane wave (\ref{e}),
\label{lemma3}
\begin{eqnarray}
\lefteqn{ -\sum_{n=0}^\infty \frac{1}{n!} \int_0^1 (\alpha-\alpha^2)^n 
\: (\OBox^n V)_{|\alpha y + (1-\alpha) x} \:d\alpha \;
\left(\frac{d}{da}\right)^{n+1} \left( a \:\log(a) \:J_a(p) \right) }
        \label{4e}  \\
         & = &-\frac{1}{2} \:\frac{d}{da} \int_{-1}^1 d\mu \:
\left( (a+b) \: \log(a+b) \int_0^1 L_{\tau a + (\tau-1) b - \mu 
pq}(p) \: d\tau \right)_{|a=m^2-\frac{q^2}{4},\;
b=\mu^2 \frac{q^2}{4}} \;\;\; . \spc
        \label{4f}
\end{eqnarray}
\end{Lemma}
{\Proof}
The series (\ref{4e}) is obtained from the formula (\ref{13}) for $A(x,y)$, 
by the replacement $T_a \rightarrow a \log(a) J_a$. As 
remarked in the proof of Theorem \ref{thm1}, all the transformations 
from (\ref{31p}) to (\ref{3e}) are also valid if we replace $T_a$ by 
any other function. Therefore,
\begin{eqnarray*}
(\ref{4e}) &=&-\sum_{n=0}^\infty \frac{(-1)^n}{n!} \left( \frac{q^2}{4} \right)^n
\sum_{k=0}^\infty \frac{1}{(2k+1)!} \sum_{l=0}^k
\left[\!\! \begin{array}{c} 2k \\ l \end{array} \!\!\right]
\left( \frac{q^2}{2} \right)^l \\
&&\hspace*{3cm} \times
\left(\frac{q^j}{2} \frac{\partial}{\partial p^j} \right)^{2k-2l}
\left(\frac{d}{da} \right)^{n+1+l} \left( a \:\log(a) \: J_a(p) \right)
\;\;\; .
\end{eqnarray*}
We carry out the sum over $n$ by redefining $a$ as $a=m^2-q^2/4$ and
substitute (\ref{4c}) as follows:
\begin{eqnarray*}
(\ref{4e}) &=&-\sum_{k=0}^\infty \frac{1}{(2k+1)!} \sum_{l=0}^k
\left[\!\! \begin{array}{c} 2k \\ l \end{array} \!\!\right]
\left( \frac{q^2}{2} \right)^l \\
&&\hspace*{2cm} \times \;
\left(\frac{q^j}{2} \frac{\partial}{\partial p^j} \right)^{2k-2l}
\left(\frac{d}{da} \right)^{1+l} \left( a \:\log(a) \: \int_0^1 
L_{\tau a}(p) \:d\tau \right) \;\;\; .
\end{eqnarray*}
Using that $L_a(p)$ and $T_a(p)$ obey the analogous
differentiation rules (\ref{4d}) and (\ref{16}), respectively, we can apply 
relation (\ref{20}) with $T_a^{(r)}$ replaced by $L_a^{(r)}$ to obtain
\begin{eqnarray*}
(\ref{4e}) &=& -\sum_{k=0}^\infty \frac{1}{(2k+1)!} \sum_{l=0}^{k}
\left[\!\! \begin{array}{c} 2k \\ l \end{array} \!\!\right]
\sum_{s=0}^{k-l} (-1)^s \left[\!\! \begin{array}{c} 2k-2l \\ s \end{array} \!\!\right]
\left( \frac{q^2}{2} \right)^{l+s} \\
&&\hspace*{2cm} \times
(pq)^{2k-2l-2s} \left(\frac{d}{da} \right)^{1+l} \left( a \:\log(a) \:
\int_0^1 L^{(2k-2l-s)}_{\tau a}(p) \:d\tau \right) \;\;\; .
\end{eqnarray*}
We introduce the index $r=l+s$, replace $s$ by $r-l$, and substitute 
the combinatorial formula (\ref{25z}):
\begin{eqnarray*}
(\ref{4e}) &=& -\sum_{k=0}^\infty \frac{(-1)^r}{2k+1} \sum_{r=0}^k
\frac{1}{r! \: (2k-2r)!} \left( \frac{q^2}{4} \right)^r
(pq)^{2k-2r} \\
&&\hspace*{2cm} \times \sum_{l=0}^r (-1)^l
\left(\!\! \begin{array}{c} r \\ l \end{array} \!\!\right)
\left(\frac{d}{da} \right)^{1+l} \left( a \:\log(a) \:
\int_0^1 L^{(2k-2r+(r-l))}_{\tau a}(p) \:d\tau \right) \;\;\; .
\end{eqnarray*}
The last sum can be eliminated using the combinatorics of the product 
rule,
\begin{eqnarray*}
\sum_{l=0}^r (-1)^l
\lefteqn{ \left(\!\! \begin{array}{c} r \\ l \end{array} \!\!\right)
\left(\frac{d}{da} \right)^{l} \left( a \:\log(a) \:
L^{(2k-2r+(r-l))}_{\tau a}(p) \right) } \\
&=& (-1)^r \left(\frac{d}{db} \right)^r \left( (a+b) \: \log(a+b) \:
L_{\tau (a+b) - b}^{(2k-2r)}(p) \right)_{|b=0} \;\;\; .
\end{eqnarray*}
Furthermore, we shift the index $k$ according to $k-r \rightarrow k$, yielding
\begin{eqnarray*}
(\ref{4e}) &=& -\sum_{k=0}^\infty \frac{1}{2k+2r+1} \frac{(pq)^{2k}}{(2k)!}
\sum_{r=0}^\infty \frac{1}{r!} \left( \frac{q^2}{4} \right)^r \\
&&\hspace*{1cm} \times
\left(\frac{d}{da} \right) \left(\frac{d}{db} \right)^r
\left( (a+b) \: \log(a+b) \int_0^1 L_{\tau a + (\tau-1) b}^{(2k)}(p) 
\: d\tau \right)_{|b=0} \;\;\; .
\end{eqnarray*}
Without the factor $(2k+2r+1)^{-1}$, we had two separate 
Taylor series which could easily be carried out 
explicitly. The coupling of the two series by this factor can be 
described with an additional line integral,
\begin{eqnarray*}
(\ref{4e}) &=&-\frac{1}{2} \:\frac{d}{da} \int_{-1}^1 d\mu \:
\sum_{r=0}^\infty \frac{1}{r!} \left( \frac{q^2}{4} 
\right)^r \mu^{2r} \\
&& \hspace*{1cm}\times\;
\left( \frac{d}{db} \right)^r \left( (a+b) \: \log(a+b) \int_0^1
L_{\tau a + (\tau-1) b + \mu pq}(p) \:d\tau \right)_{|b=0} \;\;\; .
\end{eqnarray*}
We finally carry out the remaining Taylor sum.
\QED
The result of this lemma is quite complicated. The important point is 
that the convergence problems of the infinite series (\ref{4e}) have 
disappeared in (\ref{4f}), which is obviously finite. Namely, the 
$a$-derivative of the integrand in (\ref{4f}) has at most logarithmic 
singularities. These singularities are integrable and disappear when 
the $\mu$-integration is carried out.

After these preparations, we can state the main theorem.
Since the resulting expansion is regular in the limit $m \rightarrow 
0$, it is also valid for $m=0$.
\begin{Thm} {\bf{(light-cone expansion of $\Delta T_{m^2}$)}}
\label{thm2}
The distribution $\Delta T_{m^2}$ of (\ref{7}) has the representation
\begin{eqnarray}
        \Delta T_{m^2}(x,y) & = & -\sum_{n=0}^\infty \frac{1}{n!} \:
        \int_0^1 (\alpha-\alpha^2)^n \: (\OBox^n V)_{|\alpha y+(1-\alpha)x } 
        \: d\alpha \; T^{\mbox{\scriptsize{reg }} (n+1)}_{m^2}(x,y)
        \label{4g}  \\
         &  & +N_{m^2}(x,y)
        \label{4h}
\end{eqnarray}
with
\[ T^{\mbox{\scriptsize{reg}}}_a \;=\; T_a \:-\: a \:\log(a) \:J_a 
\;\;\;,\spc
T^{\mbox{\scriptsize{reg }}(n)}_a \;=\; \left(\frac{d}{da}\right)
T^{\mbox{\scriptsize{reg}}}_a \]
and the Bessel function $J_a$ of (\ref{4a}).
The series (\ref{4g}) is well-defined in the sense of Def.\ \ref{def1}.
The contribution $N_{m^2}(x,y)$ is a smooth function in $x, y$ and has 
a representation as the Fourier integral
\begin{eqnarray}
        N_{m^2}(x,y) & = & \int \frac{d^4p}{(2 \pi)^4} \int \frac{d^4q}{(2 
        \pi)^4} \:\tilde{V}(q) \:
        N_{m^2}(p,q) \: e^{-ip(x-y)} \: e^{-i\frac{q}{2} \:(x+y)}
        \spc {\mbox{with}} \spc
        \label{4z} \\
        N_a(p,q) & = & -\frac{1}{2} \:\frac{d}{da} \int_{-1}^1 d\mu \:
        \log \left( a-(1-\mu^2) \:\frac{q^2}{4} \right) \; \left. \beta \:J_\beta(p)
        \right|^{\beta=a-\frac{q^2}{4}+\mu \:pq}_{\beta=-\mu^2 
        \frac{q^2}{4} + \mu \:pq} \;\;\; .
        \label{4k}
\end{eqnarray}
\end{Thm}
{\Proof}
By definition, $T^{\mbox{\scriptsize{reg}}}_a$ differs from $T_a$ by 
the contribution (\ref{46}). Thus an explicit formula for
$T^{\mbox{\scriptsize{reg}}}_a$ is obtained from (\ref{10}) if
we replace the factor $\log(a \xi^2- i \varepsilon \xi^0)$ in the second
line by $\log(\xi^2-i\varepsilon \xi^0)$. As a consequence,
$T^{\mbox{\scriptsize{reg}}}_a$ is a power series in $a$, and 
the higher order contributions in $a$ are of higher order on the 
light cone. This justifies the infinite sum in (\ref{4g}) in the 
sense of Def.\ \ref{def1}. Furthermore, the difference between the formal 
light-cone expansions (\ref{22}) and (\ref{4g}) coincides with the contribution 
(\ref{4e}), which was resummed in Lemma \ref{lemma3}. We carry out 
the $\tau$-integration in (\ref{4f}) using the series expansions (\ref{4a})
and (\ref{4b}), which gives (\ref{4k}).

The $q$-integral in (\ref{4z}) is well-defined since $\tilde{V}(q)$ 
is $C^1$ and decays sufficiently fast at infinity. Finally, the 
$p$-integration can be carried out with the $\delta^4$-distributions 
in (\ref{4a}), which gives a smooth function $N_{m^2}(x,y)$.
\QED
We call (\ref{4g}) and (\ref{4h}) the {\em{causal}} and {\em{noncausal 
contributions}}, respectively.

We could proceed by studying the noncausal contribution more 
explicitly in position space. For the purpose of this paper, 
however, it is sufficient to notice that $N(x,y)$ is smooth on the 
light cone.

\section{The Light-Cone Expansion of the Dirac Sea}
\setcounter{equation}{0}
Having performed the light-cone expansion for $\Delta T_{m^2}$, we now
return to the study of the Dirac sea (\ref{4}). From 
the theoretical point of view, the light-cone expansion for $\Delta 
P_{m^2}$ is an immediate consequence of Theorem \ref{thm2} and 
formula (\ref{6}): We substitute the light-cone expansion 
(\ref{4g})--(\ref{4h}) into (\ref{6}). Calculating the partial derivatives 
$\partial_x$ and $\partial_y$ of the causal contribution (\ref{4g})
gives expressions of the form
\begin{equation}
\Delta P_{m^2}(x,y) \;\asymp\;
\int_0^1 {\cal{P}}(\alpha) \: D^a \OBox^b V_{|\lambda y + 
        (1-\lambda) x} \:d\alpha \; D^c T^{\mbox{\scriptsize{reg 
        }}(n+1)}_{m^2}(x,y) \;\;\; ,
        \label{51}
\end{equation}
which are again causal in the sense that they depend on the potential and its
partial derivatives only along the line segment $\overline{xy}$
(here ${\cal{P}}(\alpha)$ denotes a polynomial in $\alpha$; $D^a$ stands 
for any partial derivatives of the order $a$).
Since (\ref{51}) contains distributional derivatives of
$T^{\mbox{\scriptsize{reg}} (n+1)}_{m^2}$, it is in general more 
singular on the light cone than the corresponding contribution to 
$\Delta T_{m^2}$. On the other hand, the partial derivatives of the noncausal 
contribution $N_{m^2}(x,y)$ can be calculated 
with (\ref{4k}) and yield smooth functions in $x, y$. We conclude
that the qualitative picture of Theorem \ref{thm2}, especially the 
splitting into a causal and a noncausal contributions, is also valid for 
the Dirac sea.

The situation becomes more complicated if one wants to go beyond 
this qualitative picture and is interested in explicit formulas for 
the Dirac sea. The problem is to find an effective and reliable method 
for calculating the partial derivatives and for handling the combinatorics 
of the Dirac matrices. Before entering these 
computational details, we explain how the qualitative picture of the 
light-cone expansion can be understood directly from the integral 
formula (\ref{4}):
The tempered distributions $s(x,y)$ and $P(x,y)$ are regular functions 
for $(y-x)^2 \neq 0$ and are singular on the light cone (this can be 
seen explicitly from e.g.\ (\ref{9a}) and (\ref{10})). Integrals of 
the form
\[ \int P(x,z) \: f(z) \: d^4z \spc {\mbox{or}} \spc
\int s(x,z) \: f(z) \: d^4z \]
with a smooth function $f$ (which decays sufficiently fast at 
infinity) give smooth functions in $x$.
The integral in (\ref{4}) is more complicated because it contains
two distributional factors $s, P$. This causes complications only if the 
singularities of $s$ and $P$ meet; i.e., if $z$ lies on the intersection 
$L_x \cap L_y$ of the light cones around $x, y$, where
\[ L_x \;=\; \left\{ y \in \R^4 \:,\; (y-x)^2=0 \right\} \;\;\; . \]
If $y-x$ is timelike or spacelike, then $L_x \cap L_y$ is a 2-sphere 
or a hyperboloid (respectively), either of which depends smoothly on
$x, y$. As a
consequence, the integral over these singularities can be carried out 
in (\ref{4}) and gives a smooth function. On the light cone 
$(y-x)^2=0$, however, $L_x \cap L_y$ does not depend smoothly on $x, y$. 
More precisely, in the limit $0<(y-x)^2 \rightarrow 0$, the 2-sphere 
$L_x \cap L_y$ degenerates to the line segment
$\{ \lambda y + (1-\lambda x) \:,\; 0 \leq \lambda \leq 1\}$.
The limit $0>(y-x)^2 \rightarrow 0$, on the other hand, gives the degenerated
hyperboloid
$\{ \lambda y + (1-\lambda x)\:,\; \lambda \leq 0 {\mbox{ or }} 
\lambda \geq 1\}$.
This simple consideration explains why the singularities of $\Delta 
P(x,y)$ occur on the light cone and makes it plausible that the 
behavior of the singularities is characterized by the potential and 
its derivatives along the line $xy=\{ \lambda y + (1-\lambda) x\:,\; 
\lambda \in \R\}$. Clearly, $V(z)$ also enters into $\Delta 
P_{m^2}(x,y)$ for $z \not \in xy$, but this noncausal 
contribution is not related to the discontinuity of $L_x \cap L_y$ on 
the light cone and is therefore smooth. The special form of the singularities,
\begin{equation}
        \Delta P(x,y) \;\sim\; D^a \log((y-x)^2- i 0 (y-x)^0) \; (y-x)^{2n}
        \;\;\; ,
        \label{52}
\end{equation}
is less obvious.
That the potential enters only along the line segment
$\overline{xy}$ can be understood only from the special form of 
(\ref{4}); it is a consequence of the causality principle for the 
Dirac sea that was introduced in \cite{F1}.
In fact, it gives an easy way to understand the meaning of 
``causality'' of the perturbation expansion for the Dirac sea.

We finally describe our method for explicitly calculating $\Delta 
P(x,y)$. As in Theorem \ref{thm2}, we will not study the 
noncausal contribution; we are content with the fact that it is bounded and 
smooth. In other words, we consider only the singular contribution (\ref{51})
to the Dirac sea. Since the difference between $T_{m^2}^{(n)}$ and 
$T_{m^2}^{\mbox{\scriptsize{reg }}(n)}$ is smooth, we can just as well 
consider the formal light-cone expansion (\ref{22}) and calculate 
modulo smooth terms on the light cone. This has the advantage that we 
can work with the useful differentiation rule (\ref{11}).
The calculation can be split into several steps, which may be listed as
follows.
\begin{description}
\item[(1)] {\em{Calculation of the partial derivatives}} with the product rule 
and the differentiation formulas
\begin{eqnarray}
\lefteqn{ \hspace*{-4.8cm}
\frac{\partial}{\partial x^j} T_{m^2}^{(n)}(x,y) \;=\;
        -\frac{\partial}{\partial y^j} T_{m^2}^{(n)}(x,y)
        \;\stackrel{(\ref{11})}{=}\;
        \frac{1}{2} \:(y-x)_j \; T_{m^2}^{(n-1)}(x,y) \spc
        \label{54} } \\
        \frac{\partial}{\partial y^j} \int_0^1 {\cal{P}}(\alpha) \:
        D^a \OBox^b V_{|\alpha y + (1-\alpha) x} & = & 
        \int_0^1 \alpha \:{\cal{P}}(\alpha) \:
        \partial_j D^a \OBox^b V_{|\alpha y + (1-\alpha) x} \\
        \frac{\partial}{\partial x^j} \int_0^1 {\cal{P}}(\alpha) \:
        D^a \OBox^b V_{|\alpha y + (1-\alpha) x} & = & 
        \int_0^1 (1-\alpha) \:{\cal{P}}(\alpha) \:
        \partial_j D^a \OBox^b V_{|\alpha y + (1-\alpha) x} \\
        \frac{\partial}{\partial x^j} (y-x)_k & = &
        -\frac{\partial}{\partial y^j} (y-x)_k \;=\; -g_{jk} \;\;\; .
\end{eqnarray}
\item[(2)] {\em{Simplification of the Dirac matrices}} with the
anti-commutation relations $\{\gamma^j,\:\gamma^k\}=2g^{jk}$. This leads to a 
contraction of tensor indices. The generated factors $(y-x)^2$ and 
$(y-x)^j \partial_j V$ are simplified in the following calculation
steps {\bf{(3)}} and {\bf{(4)}}.
\item[(3)] {\em{Absorption of the factors $(y-x)^2$}}. 
We calculate the Laplacian by iterating (\ref{54}),
\[ \OBox_x T_{m^2}^{(n+2)}(x,y) \;=\; -2 T_{m^2}^{(n+1)}(x,y) \:+\:
\frac{1}{4} \:(y-x)^2 \:T_{m^2}^{(n)}(x,y) \;\;\; , \]
and then combine it with (\ref{12}), which gives the rule
\begin{equation}
        (y-x)^2 \:T_{m^2}^{(n)}(x,y) \;=\; -4n T_{m^2}^{(n+1)}(x,y) \:-\: 4 m^2 \: 
        T_{m^2}^{(n+2)}(x,y) \;\;\; .
\end{equation}
\item[(4)] {\em{Partial integration of the tangential derivatives}},
\begin{eqnarray}
\lefteqn{ \int_0^1 {\cal{P}}(\alpha) \: (y-x)^j \partial_j D^a 
\OBox^b V_{|\alpha y + (1-\alpha) x} \;=\;
\int_0^1 {\cal{P}}(\alpha) \: \frac{d}{d\alpha} D^a 
\OBox^b V_{|\alpha y + (1-\alpha) x} } \nonumber \\
         & = & \left. {\cal{P}}(\alpha) \: D^a \OBox^b V_{|\alpha y + (1-\alpha) x}
         \right|^{\alpha=1}_{\alpha=0} \:-\: \int_0^1 {\cal{P}}^\prime(\alpha)
         \: D^a \OBox^b V_{|\alpha y + (1-\alpha) x} \;\;\; .
\end{eqnarray}
\end{description}
After these steps, $\Delta P_{m^2}(x,y)$ consists of many terms of the 
form
\[ \Delta P_{m^2}(x,y) \;\asymp\;
{\mbox{\em{(causal expression in $D^a \OBox^b V$)}}} \:\times\: 
T_{m^2}^{(n)}(x,y) \;\;\;,\;\;\;\;\;n \geq -1 \;\;\; . \]
It remains to insert the series representations for 
$T_{m^2}^{(n)}(x,y)$. It is useful first to introduce the short notation
$z^n=\xi^{2n}$ ($n \geq 0$) and
\begin{eqnarray*}
        z^{-2} & := & \frac{1}{2} \lim_{0<\varepsilon \rightarrow 0} \sum_{\pm}
        \frac{1}{(\xi^2- i \varepsilon \xi^0)^2}  \\
        z^{-1} & := & \frac{1}{2} \lim_{0<\varepsilon \rightarrow 0} \sum_{\pm}
        \frac{1}{\xi^2- i \varepsilon \xi^0}  \\
        \log z & := & \frac{1}{2} \lim_{0<\varepsilon \rightarrow 0} \sum_{\pm}
        \log( \xi^2 - i \varepsilon \xi^0) \;\;\; .
\end{eqnarray*}
\begin{description}
        \item[(5)]  {\em{Substitution of the explicit formulas}}
\begin{eqnarray}
        T_{m^2}^{(-1)}(x,y) &
        \asymp&-\frac{1}{2 \pi^3} \: z^{-2}
        \:-\: \frac{m^2}{8 \pi^3} \: z^{-1}
        \:-\: \frac{1}{8 \pi^3} \:\sum_{l=0}^\infty \frac{(-1)^{l+1}}{4^{l+1}}
        \:\frac{m^{2l+4}}{l!\:(l+2)!} \: z^l \: \log z \spc \\
        T_{m^2}(x,y) &\asymp& -\frac{1}{8 \pi^3} \: z^{-1}
        \:-\: \frac{1}{8 \pi^3} \:\sum_{l=0}^\infty \frac{(-1)^{l+1}}{4^{l+1}}
        \:\frac{m^{2l+2}}{l!\:(l+1)!} \: z^l \: \log z \\
        T_{m^2}^{(n)}(x,y) &\asymp&-\frac{1}{8 \pi^3} \:
        \sum_{l=n-1}^\infty \frac{(-1)^{l+1}}{4^{l+1}}
        \:\frac{m^{2l+2-2n}}{l!\:(l+1-n)!} \: z^l \:\log z \;\;\; , \;\;\;\;\;
        (n\geq1), \label{55}
\end{eqnarray}
where again we have used the notation of (\ref{10}) (we take only the singular
contribution on the light cone; $T_{m^2}^{(-1)}$ is defined via (\ref{54})).
\end{description}
In this way, the calculation of the causal contribution is reduced to a 
small number of symbolic computation rules (\ref{54})--(\ref{55}), 
which can be applied mechanically. This makes it possible to use a 
computer program for the calculation. The C++ program 
``class{\_}commute'' was designed for this task (commented 
source code available from the
author on request). It computes the causal contribution for a general 
perturbation (\ref{00}) to any order on the light cone. The formulas 
to the order ${\cal{O}}((y-x)^0)$ modulo the noncausal contribution
are listed in the appendix.

\section{Outlook}
In this paper, the light-cone expansion was performed for the 
Dirac sea to first order in the external potential. The presented 
method can be generalized in several directions and applied to
related problems, which we now briefly outline.

First of all, the method is not restricted to the Dirac and 
Klein-Gordon equations; it can also be used for the analysis of other
scalar and matrix hyperbolic equations (in any space-time 
dimension). The consideration (\ref{0a}), which gives the basic 
explanation for the line integrals in the light-cone expansion, can be 
applied to any hyperbolic equation (in curved space-time, the line 
integrals must be replaced by integrals along null geodesics; see e.g.\
\cite{FL}). Thus, the behavior of the solution near the light cone is 
again described by an infinite series of line integrals. The line 
integrals might be unbounded, however, which leads to additional 
convergence problems (e.g., one can replace the integrals in (\ref{13})
by $\frac{1}{2} \int_{-\infty}^\infty \epsilon(\alpha) \:d\alpha 
\cdots$, which gives a different formal solution of (\ref{14})).

Furthermore, the light-cone expansion can be generalized to symmetric 
eigensolutions and the fundamental solutions:
The formal light-cone expansion of 
Section 3 applies in the same way to any Lorentzian invariant 
family $T_{m^2}$ of solutions of the Klein-Gordon equation, i.e.\
to a linear combination of
\begin{eqnarray}
        T_{m^2}(p) & = & \delta(p^2-m^2) \spc\;\;\;\; {\mbox{and}}
        \label{61}  \\
        T_{m^2}(p) & = & \delta(p^2-m^2) \:\epsilon(p^0) \;\;\; .
        \label{62}
\end{eqnarray}
The second case (\ref{62}) allows us to generalize the light-cone expansion
to the Green's function. The advanced
Green's function $S^{\vee}_{m^2}$ of the Klein-Gordon operator, for example,
can be derived from $T_{m^2}$ of (\ref{62}) by
\begin{eqnarray*}
        S^{\vee}_{m^2}(x,y) & = & 2 \pi i \: T_{m^2}(x,y) \: \Theta(y^0-x^0) 
        \;\;\; .
\end{eqnarray*}
This relation even remains valid in the perturbation expansion, e.g.\
to first order,
\begin{eqnarray}
        \Delta S^{\vee}_{m^2}(x,y) & = & 2 \pi i \: \Delta T_{m^2}(x,y) \:
        \Theta(y^0-x^0)
        \label{63} 
\end{eqnarray}
(for a derivation of this formula in the context of the Dirac equation, see 
\cite{F1}).
Thus the light-cone expansion for $\Delta T_{m^2}$ immediately yields 
corresponding formulas for the Green's function.

In contrast to the formal light-cone expansion of Section 2, the 
resummation of the noncausal contribution depends much on the 
particular problem. An analysis in position space according to 
\cite{F2} might be helpful for the understanding of the noncausality. 
For $T_{m^2}$ according to (\ref{62}), for example, there is no 
noncausal contribution at all, which also simplifies the analysis of the 
Green's functions.

By iteration, the method can also be applied to higher-order Feynman diagrams
and even makes it possible to sum up certain classes of Feynman 
diagrams explicitly. For the Dirac Green's function and the Dirac sea,
this is explained in detail in \cite{F3}.

\appendix
\section{Some Formulas of the Light-Cone Expansion}
\label{app_A}
\setcounter{equation}{0}
The following formulas give $\Delta P(x,y)$ to first order in the 
external potential (\ref{00}) up to contributions of the order
${\cal{O}}((y-x)^0)$ on the light cone. For the causal line integrals, we
use the short notation
\[ \int_x^y f \cdots \;:=\; \int_0^1 f_{|\alpha y + (1-\alpha) x} 
\:\cdots\: \:d\alpha \;\;\; . \]
\subsection{Electromagnetic Potential}
\begin{eqnarray*}
\Delta P(x,y) &=& - \frac{e}{4 \pi^3}
\:\int_x^y A_j \: \xi^j \; \xi\slsh \: z^{-2} \\
&& - \frac{e}{16 \pi^3} \:\int_x^y (\alpha^2-\alpha) \; \xi \slsh
        \: \xi^k \: j_k \; z^{-1} \\
&& + \frac{e}{16 \pi^3} \: \int_x^y (2 \alpha - 1) \; \xi^j \:
        \gamma^k \: F_{kj} \;z^{-1} \\
&& + \frac{i e}{32 \pi^3} \:\int_x^y \varepsilon^{ijkl} \; F_{ij} \:
        \xi_k \; \gamma^5 \gamma_l \; z^{-1} \\
&& - \frac{e}{128 \pi^3} \: \int_x^y (\alpha^4 - 2 \alpha^3
        + \alpha^2) \: \xi \slsh \: \xi_k \; \OBox j^k \; \log z \\
&& + \frac{e}{128 \pi^3} \: \int_x^y (4 \alpha^3
        - 6 \alpha^2 + 2 \alpha) \: \xi^j \: \gamma^k \: (\OBox F_{kj})
        \;\log z \\
&& + \frac{i e}{128 \pi^3} \: \int_x^y (\alpha^2 - \alpha) \;
        \varepsilon^{ijkl} \: (\OBox F_{ij}) \: \xi_k \; \gamma^5 \gamma_l
        \;\log z \\
&& + \frac{e}{16 \pi^3} \: \int_x^y (\alpha^2 - \alpha)
        \: \gamma^k \: j_k \;\; \log z \\
&&+\frac{ie}{8 \pi^3} \;m\: \int_x^y A_j \xi^j \: z^{-1} \\
&& - \frac{e}{64 \pi^3} \; m \; \int_x^y F_{ij} \:
        \sigma^{ij} \: \log z \\
&& - \frac{ie}{32 \pi^3} \; m \: \int_x^y (\alpha^2-\alpha) \:
        j_k \: \xi^k \: \log z \\
&&-\frac{e}{16 \pi^3} \; m^2 \: \int_x^y A_j \xi^j \; \xi\slsh \:z^{-1} \\
&& - \frac{e}{64 \pi^3} \; m^2 \: \int_x^y (2 \alpha -1) \:
        \gamma^i \: F_{ij} \: \xi^j \: \log z \\
&& - \frac{ie}{128 \pi^3} \; m^2 \: \int_x^y \varepsilon^{ijkl}
        \; F_{ij} \:  \xi_k \; \gamma^5 \gamma_l \: \log z \\
&&+\frac{e}{64 \pi^3} \:m^2\: \int_x^y (\alpha^2-\alpha) \: j_k \:
\xi^k \: \xi\slsh \: \log z \\
&&-\frac{ie}{32 \pi^3} \; m^3 \:\int_x^y A_j \xi^j \: \log z \\
&&+\frac{e}{128 \pi^3} \:m^4\:\int_x^y A_j \xi^j \:\xi\slsh\:\log z \\
&&+ {\mbox{(noncausal contributions)}} \:+\: {\cal{O}}(\xi^2)
\end{eqnarray*}
with the electromagnetic field tensor $F_{jk}=\partial_j A_k - \partial_k A_j$
and the electromagnetic current $j^k=\partial_l F^{kl}$.

\subsection{Axial Potential}
\begin{eqnarray*}
\Delta P(x,y) &=& \frac{e}{4 \pi^3}
\: \int_x^y B_j \: \xi^j \; \gamma^5 \xi\slsh \: z^{-2} \\
&& +\frac{e}{16 \pi^3} \:\int_x^y (\alpha^2-\alpha) \; \gamma^5 \xi \slsh
        \: \xi^k \: j_k \; z^{-1} \\
&& - \frac{e}{16 \pi^3} \:\int_x^y (2 \alpha - 1) \; \xi^j \:
        \gamma^5 \gamma^k \: F_{kj} \;z^{-1} \\
&& - \frac{i e}{32 \pi^3} \:\int_x^y \varepsilon^{ijkl} \; F_{ij} \:
        \xi_k \; \gamma_l\; z^{-1} \\
&& + \frac{e}{128 \pi^3} \: \int_x^y (\alpha^4 - 2 \alpha^3
        + \alpha^2) \: \gamma^5 \xi \slsh \: \xi_k \; \OBox j^k \; \log z \\
&& - \frac{e}{128 \pi^3} \: \int_x^y (4 \alpha^3
        - 6 \alpha^2 + 2 \alpha) \: \xi^j \: \gamma^5 \gamma^k \: (\OBox F_{kj})
        \;\log z \\
&& - \frac{i e}{128 \pi^3} \: \int_x^y (\alpha^2 - \alpha) \;
        \varepsilon^{ijkl} \: (\OBox F_{ij}) \: \xi_k \; \gamma_l
        \;\log z \\
&& - \frac{e}{16 \pi^3} \: \int_x^y (\alpha^2 - \alpha)
        \: \gamma^5 \gamma^k \: j_k \;\; \log z \\
&& - \frac{ie}{8 \pi^3} \: m \: \int_x^y \gamma^5 \:
        \frac{1}{2} [\xi\slsh, B \!\slsh] \: z^{-1} \\
&& -\frac{e}{64 \pi^3} \: m \: \int_x^y (2\alpha -1) \;
        F_{jk} \; \gamma^5 \sigma^{jk} \: \log z \\
&& +\frac{ie}{32 \pi^3} \: m \: \int_x^y \partial_j B^j \;
        \gamma^5 \: \log z \\
&& +\frac{e}{32 \pi^3} \: m \:\int_x^y (\alpha^2-\alpha) \;
        \OBox B_j \; \xi_k \; \gamma^5 \sigma^{jk} \: \log z \\
&& +\frac{e}{16 \pi^3} \: m^2 \: \int_x^y B_j \: \xi^j \;
        \gamma^5 \xi\slsh \: z^{-1} \\
&& -\frac{e}{16 \pi^3} \: m^2 \:\int_x^y \gamma^5 B\!\slsh \: \log z \\
&& +\frac{e}{64 \pi^3} \: m^2 \:\int_x^y (2\alpha-1) \;
        F_{jk} \: \xi^k \; \gamma^5 \gamma^j \: \log z \\
&& +\frac{ie}{128 \pi^3} \: m^2 \:\int_x^y \varepsilon^{ijkl}
        \; F_{ij} \: \xi_k \; \gamma_l \: \log z \\
&& -\frac{e}{64 \pi^3} \: m^2 \:\int_x^y (\alpha^2-\alpha) \;
        j_k \: \xi^k \; \gamma^5 \xi\slsh \: \log z \\
&& +\frac{ie}{32 \pi^3} \: m^3 \:\int_x^y \gamma^5\:
        \frac{1}{2} [\xi\slsh, B\!\slsh] \: \log z \\
&&-\frac{e}{128 \pi^3} \:m^4 \: \int_x^y A_j \xi^j \:\gamma^5 \xi\slsh
\: \log z \\
&&+ {\mbox{(noncausal contributions)}} \:+\: {\cal{O}}(\xi^2)
\end{eqnarray*}
with the axial field tensor $F_{jk}=\partial_j B_k - \partial_k B_j$
and the axial current $j^k=\partial_l F^{kl}$.

\subsection{Scalar Potential}
\begin{eqnarray*}
\Delta P(x,y) &=& \frac{1}{16 \pi^3} \: (\Phi(y)+\Phi(x)) \; z^{-1} \\
&&+ \frac{i}{16 \pi^3} \: \int_x^y
        \; (\partial_j \Phi) \; \xi_k \; \sigma^{jk} \: z^{-1} \\
&&+ \frac{i}{64 \pi^3} \: \int_x^y
        (\alpha^2 - \alpha) \; (\partial_j \OBox \Phi) \;
        \xi_k \; \sigma^{jk} \: \log z \\
&&- \frac{1}{64 \pi^3} \: \int_x^y \OBox \Phi \: \log z \\
&&+ \frac{i}{8 \pi^3}\: m \: \int_x^y \Phi \; \xi\slsh \: z^{-1} \\
&&+ \frac{i}{32 \pi^3} \: m \: \int_x^y (2\alpha-1) \;
        (\Pdd \Phi) \: \log z \\
&&+ \frac{i}{32 \pi^3} \: m \: \int_x^y (\alpha^2-\alpha) \;
        (\OBox \Phi) \; \xi\slsh \: \log z \\
&&- \frac{1}{64 \pi^3} \: m^2 \:  (\Phi(y) + \Phi(x)) \: \log z \\
&&- \frac{1}{16 \pi^3} \: m^2 \: \int_x^y \Phi \: \log z \\
&&- \frac{i}{64 \pi^3} \: m^2 \: \int_x^y 
        (\partial_j \Phi) \: \xi_k \; \sigma^{jk} \: \log z \\
&&-\frac{i}{32 \pi^3} \:m^3\: \int_x^y \Phi \:\xi\slsh\: \log z \\
&&+ {\mbox{(noncausal contributions)}} \:+\: {\cal{O}}(\xi^2)
\end{eqnarray*}

\subsection{Pseudoscalar Potential}
\begin{eqnarray*}
\Delta P(x,y) &=&  -\frac{i}{16 \pi^3} \: (\Xi(y)+\Xi(x)) \;\gamma^5 \:
 z^{-1} \\
&&+ \frac{1}{16 \pi^3} \int_x^y
        \; (\partial_j \Xi) \; \xi_k \; \gamma^5 \sigma^{jk} \: z^{-1} \\
&&+ \frac{1}{64 \pi^3} \: \int_x^y
        (\alpha^2 - \alpha) \; (\partial_j \OBox \Xi) \;
        \xi_k \; \gamma^5 \sigma^{jk} \: \log z \\
&&+ \frac{i}{64 \pi^3} \: \int_x^y
        \OBox \Xi \: \gamma^5 \: \log z \\
&&- \frac{1}{32 \pi^3} \: m \: \gamma^5 \: \int_x^y (\Pdd \Xi) \: \log z \\
&&+ \frac{i}{64 \pi^3} \: m^2 \: (\Xi(y) + \Xi(x)) \; \gamma^5 \: \log z \\
&&- \frac{1}{64 \pi^3} \: m^2 \int_x^y 
        (\partial_j \Xi) \: \xi_k \; \gamma^5 \sigma^{jk} \: \log z \\
&&+ {\mbox{(noncausal contributions)}} \:+\: {\cal{O}}(\xi^2)
\end{eqnarray*}

\subsection{Bilinear Potential}
\begin{eqnarray*}
\label{eq:a1_b11}
\Delta P(x,y) &=& -\frac{1}{2\pi^3} \;
        \int_x^y H_{ij} \; \xi^i \: \xi_k \; \sigma^{jk} \: z^{-2}\\
&&+ \frac{1}{16 \pi^3} \; (H_{jk}(y) +
        H_{jk}(x)) \; \sigma^{jk} \: z^{-1} \\
&&- \frac{1}{4 \pi^3} \;\int_x^y H_{jk} \;
        \sigma^{jk} \: z^{-1} \\
&&+ \frac{i}{8 \pi^3} \;\int_x^y \xi_j \;
        H^{jk}_{\;\;,k} \: z^{-1} \\
&&+ \frac{1}{8 \pi^3} \;\int_x^y (2\alpha-1) \;
        (\xi^k \: H_{jk,i} + \xi_i \: H_{jk,}^{\;\;\;k}) \;
        \sigma^{ij} \: z^{-1} \\
&&+ \frac{1}{8 \pi^3} \;\int_x^y
        (\alpha^2-\alpha) \; (\OBox H_{ij}) \: \xi^i \: \xi_k \; \sigma^{jk}
        \: z^{-1} \\
&&- \frac{1}{16 \pi^3} \;\int_x^y
        \varepsilon^{ijkl} \; H_{ij,k} \; \xi_l \; \gamma^5 \: z^{-1} \\
&&+\frac{1}{8 \pi^3} \:\int_x^y (\alpha^2-\alpha) \:\partial_j 
H_{kl,}^{\;\;\;\:l} \: \sigma^{jk} \: \log z \\
&&-\frac{1}{16 \pi^3} \:\int_x^y (\alpha^2-\alpha+\frac{1}{4}) \:
(\OBox H_{jk}) \: \sigma^{jk} \: \log z \\
&&-\frac{1}{64 \pi^3} \:\int_x^y (\alpha^2-\alpha) \: \epsilon^{ijkl}
\:(\OBox H_{ij,k}) \:\xi_l \:\gamma^5 \:\log z \\
&&+\frac{i}{32 \pi^3} \:\int_x^y (\alpha^2-\alpha) \: \xi^j \:(\OBox 
H_{jk,}^{\;\;\;\:k})\:\log z \\
&&+\frac{1}{64 \pi^3} \:\int_x^y (\alpha^2-\alpha)^2 \: (\OBox^2 
H_{jk}) \: \xi^j \:\xi_l \: \sigma^{kl} \: \log z \\
&&+\frac{1}{32 \pi^3} \: \int_x^y (2 \alpha^3-3\alpha^2+\alpha) \:
\left( \xi^k \: \OBox H_{jk,i} \:+\: \xi_i \: \OBox 
H_{jk,}^{\;\;\;\:k} \right) \sigma^{ij} \: \log z \\    
&&+ \frac{i}{8 \pi^3} \: m \: \int_x^y
        \varepsilon^{ijkl} \; H_{ij} \: \xi_k \; \gamma^5 \gamma_l \:z^{-1} \\
&&- \frac{1}{16 \pi^3} \: m \:\int_x^y
        H_{jk,}^{\;\;\;k} \; \gamma^j \: \log z \\
&&+ \frac{i}{32 \pi^3} \: m \: \int_x^y (2\alpha-1) \;
        \varepsilon^{ijkl} \: H_{ij,k} \; \gamma^5 \gamma_l \: \log z \\
&&+ \frac{i}{32 \pi^3} \: m \: \int_x^y (\alpha^2-\alpha) \;
        \varepsilon^{ijkl} \; (\OBox H_{ij}) \: \xi_k
        \; \gamma^5 \gamma_l \: \log z \\
&&- \frac{1}{8 \pi^3} \: m^2 \: \int_x^y
        H_{ij} \: \xi^i \: \xi_k \; \sigma^{jk} \: z^{-1} \\
&&- \frac{1}{64 \pi^3} \: m^2 \;
        (H_{jk}(y) + H_{jk}(x)) \; \sigma^{jk} \: \log z \\
&&- \frac{i}{32 \pi^3} \: m^2 \: \int_x^y
        \xi_j \; H^{jk}_{\;\;\:,k} \: \log z \\
&&- \frac{1}{32 \pi^3} \: m^2 \:
        \int_x^y (2\alpha-1) \; (\xi^k \: H_{jk,i} + \xi_i \:
        H_{jk,}^{\;\;\;k}) \; \sigma^{ij} \: \log z \\
&&- \frac{1}{32 \pi^3} \: m^2 \:
        \int_x^y (\alpha^2-\alpha) \; (\OBox H_{ij}) \: \xi^i \: \xi_k \;
        \sigma^{jk} \:\log z \\
&&+ \frac{1}{64 \pi^3} \: m^2 \:
        \int_x^y \varepsilon^{ijkl} \; H_{ij,k} \: \xi_l \; \gamma^5 \:
        \log z \\
&&-\frac{i}{32 \pi^3} \:m^3\: \int_x^y \epsilon^{ijkl} \:H_{ij} 
\:\xi_k \: \gamma^5 \gamma_l \: \log z \\
&&+\frac{1}{64 \pi^3} \:m^4\:\int_x^y \epsilon^{ijkl} \: H_{ij} 
\:\xi^i \xi_k \: \sigma^{jk} \: \log z \\
&&+ {\mbox{(noncausal contributions)}} \:+\: {\cal{O}}(\xi^2)
\end{eqnarray*}

\section{Perturbation by a Gravitational Field}
\label{app_b}
In this appendix, we outline how the light-cone expansion can be 
extended to a perturbation by a gravitational field. For the metric, 
we consider a perturbation $h_{jk}$ of the Minkowski metric 
$\eta_{jk}={\mbox{diag}}(1,-1,-1,-1)$,
\[ g_{jk}(x) \;=\; \eta_{jk} \:+\: h_{jk}(x) \;\;\; . \]
We describe gravitation with the linearized Einstein equations (see e.g.\ 
\cite{LL}). According to the usual formalism, we raise and lower 
tensor indices with respect to the Minkowski metric.
Using the transformation of $h_{jk}$ under infinitesimal coordinate 
transformations, we can assume \cite[Par.\ 105]{LL} that
\[ \partial^k h_{jk} \;=\; \frac{1}{2} \:\partial_j h \spc 
{\mbox{with}} \spc h \;:=\; h^k_k \;\;\; . \]
In the so-called symmetric gauge, the Dirac operator takes the form \cite{F4}
\[ i \Pdd_x \:-\: \frac{i}{2} \:\gamma^j \:h_{jk}\:\eta^{kl} 
\:\frac{\partial}{\partial x^l} \:+\: \frac{i}{8} \:(\Pdd h) \;\;\; . \]
In contrast to (\ref{1}), the perturbation is now itself a 
differential operator.

One complication arises from the fact that the integration measure in 
curved space is $\sqrt{|g|} \:d^4x = (1+\frac{h}{2}) \:d^4x$, whereas 
the formula (\ref{4}) for the perturbation of the Dirac sea is
valid only if one has the integration measure $d^4 x$ of Minkowski space. 
Therefore we first transform the system such that the integration 
measure becomes $d^4x$, then apply (\ref{4}), and finally transform back to 
the original integration measure $\sqrt{|g|} \:d^4x$. Since the 
scalar product
\[ \int \overline{\Psi} \:\Phi \:\sqrt{|g|} \:d^4x \;=\; \int 
\overline{ (|g|^{\frac{1}{4}} \Psi) } \:(|g|^{\frac{1}{4}} \:\Phi) \:d^4x \]
is coordinate-invariant, the transformation to the measure $d^4x$ is 
accomplished by
\begin{eqnarray*}
\Psi(x) & \rightarrow & \hat{\Psi}(x) \;=\;
|g|^{\frac{1}{4}}(x) \:\Psi(x)  \\
i \Pdd_x - \frac{i}{2} \:\gamma^j \:h_j^k\: \partial_k
+ \frac{i}{8} \:(\Pdd h) & \rightarrow & |g|^{\frac{1}{4}} \left(
i \Pdd_x - \frac{i}{2} \:\gamma^j \:h_j^k\: \partial_k
+ \frac{i}{8} \:(\Pdd h) \right) |g|^{-\frac{1}{4}} \\
&& \;=\; i \Pdd_x - \frac{i}{2} \:\gamma^j \:h_j^k\: \partial_k
- \frac{i}{8} \:(\Pdd h) \;\;\; .
\end{eqnarray*}
The perturbation $\Delta P^{(d^4x)}$ of the transformed system
is given by (\ref{4}),
\begin{eqnarray}
\Delta P^{(d^4x)}(x,y) &=& -\int d^4z \:
\left( s(x,z) \left( -\frac{i}{2}\:\gamma^j\:h_j^k \frac{\partial}{\partial z^k}
        - \frac{i}{8} (\Pdd h)(z) \right) P(z,y) \right. \nonumber \\
&& \hspace*{1.15cm} \left. \:+\: P(x,z) \left( -\frac{i}{2}\:\gamma^j\:h_j^k
\frac{\partial}{\partial z^k}
        - \frac{i}{8} (\Pdd h)(z) \right) s(z,y) \right) \;\;\;\;\; .
        \label{pd4x}
\end{eqnarray}
The formula for the transformation of the Dirac sea to the original 
integration measure $\sqrt{|g|} \:d^4x$ is
\[ P(x,y) \:+\: \Delta P(x,y) \;=\; |g|^{-\frac{1}{4}}(x) 
\:|g|^{-\frac{1}{4}}(y) \left( P(x,y) \:+\: \Delta P^{(d^4x)}(x,y) 
\right) \;\;\; . \]
Thus
\[ \Delta P(x,y) \;=\; \Delta P^{(d^4x)}(x,y)
\:-\:\frac{1}{4} \:( h(x) + h(y)) \:P(x,y) \;\;\; . \]
The factors $P(z,y)$ and $s(z,y)$ in (\ref{pd4x})
depend only on $(z-y)$, i.e.
\[ \frac{\partial}{\partial z^k} P(z,y) \;=\;
-\frac{\partial}{\partial y^k} P(z,y) \;\;\;,\spc
\frac{\partial}{\partial z^k} s(z,y) \;=\; 
-\frac{\partial}{\partial y^k} s(z,y) \;\;\; , \]
so we may rewrite the $z$-derivatives as $y$-derivatives, which can be 
pulled out of the integral. Furthermore, the relations
\begin{eqnarray*}
\int d^4z \; P(x,z) \:(i \Pdd_z h(z))\: s(z,y)
&=& \int d^4z \; P(x,z) \:[(i \Pdd_z-m),\: h(z)])\: s(z,y) \\
&=&-P(x,y) \:h(y) \\
\int d^4z \; s(x,z) \:(i \Pdd_z h(z))\: P(z,y) &=& h(x) \:P(x,y)
\end{eqnarray*}
allow us to simplify the factors $(\Pdd h)$ in the integral. In the 
resulting formula for $\Delta P(x,y)$, one recovers the perturbation
by an electromagnetic potential. More precisely,
\begin{equation}
\Delta P(x,y) \;=\; \left(-\frac{1}{8}\:h(x) \:-\: \frac{3}{8}\:h(y) \right)
P(x,y) \:-\: \frac{i}{2}\, \frac{\partial}{\partial y^k} \Delta P[\gamma^j 
h^k_j](x,y) \;\;\; , \label{dA}
\end{equation}
where $\Delta P[\gamma^j h^k_j](x,y)$ is the perturbation (\ref{4}) of 
the Dirac sea corresponding to the electromagnetic potential ${\cal{B}}=\gamma^j h^k_j$.
The light-cone expansion of $\Delta P(x,y)$ is obtained by substituting the
light-cone expansion of $\Delta P[\gamma^j h^k_j](x,y)$ into 
(\ref{dA}) and calculating the $y$-derivatives.
To the order ${\cal{O}}((y-x)^0)$ on the light cone, this gives the 
following formula for the light-cone expansion of the Dirac sea in the 
gravitational field:
\begin{eqnarray*}
\Delta{P}(x,y) &=& -\frac{i}{8 \pi^3}
        \left( \int_x^y h^k_j \right) \xi^j \frac{\partial}{\partial y^k}
        \;\xi\slsh \: z^{-2} \\
&&-\frac{i}{16 \pi^3} \:
        \left( \int_x^y (2 \alpha -1) \; \gamma^i \: \xi^j \:\xi^k
        \; (h_{jk},_i - h_{ik},_j)  \right) \:z^{-2}\\
&&-\frac{1}{32 \pi^3} \:
        \left( \int_x^y \varepsilon^{ijlm} \; (h_{jk},_i
        - h_{ik},_j) \: \xi^k \; \xi_l \: \rho \gamma_m \right) \:z^{-2} \\
&& + \frac{i}{16 \pi^3} \:
\left( \int_x^y (\alpha^2 - \alpha) \; \xi^j \; \xi^k \;
        R_{jk} \right) \;\xi\slsh \:z^{-2} \\
&&- \frac{i}{128 \pi^3} \:
        \left( \int_x^y (\alpha^4 - 2 \alpha^3 + \alpha^2)
        \; \xi \slsh \; \xi^j \; \xi^k \; \OBox R_{jk} \right) \:z^{-1} \\
&&+ \frac{i}{128 \pi^3}  \:
        \left( \int_x^y (6 \alpha^2 - 6 \alpha + 1) \;
        \xi \slsh \; R \right) \:z^{-1} \\
&&-\frac{i}{128 \pi^3} \:\left( \int_x^y (4 \alpha^3 - 6 \alpha^2 + 2 \alpha)
        \; \xi^j \: \xi^k \: \gamma^l \; R_{j[k},_{l]} \right) \:z^{-1} \\
&&-\frac{1}{64 \pi^3} \:\left( \int_x^y (\alpha^2 - \alpha) \;
        \varepsilon^{ijlm} \: R_{ki},_j \: \xi^k \: \xi_l \: \rho \gamma_m
        \right) \:z^{-1} \\
&&+\frac{i}{32 \pi^3} \: \left( \int_x^y (\alpha^2 - \alpha) \; \xi^j \:
        \gamma^k \: G_{jk} \right) \:z^{-1} \\
&&+ \frac{i}{32 \pi^3} \: m \left( \int_x^y h_{ki,j} \right) \: \xi^k \;
        \sigma^{ij} \: z^{-1} \\
&&+ \frac{1}{32 \pi^3} \: m \int_x^y (\alpha^2-\alpha) \;
        R_{jk} \; \xi^j \: \xi^k \; z^{-1} \\
&&- \frac{i}{64 \pi^3} \: m^2 \:\int_x^y (2\alpha-1)
        \; (h_{jk,i} - h_{ik,j}) \: \gamma^i \; \xi^j \: \xi^k \; z^{-1} \\
&&+ \frac{1}{64 \pi^3} \: m^2 \:\int_x^y \varepsilon^{ijlm}
        h_{jk,i} \; \xi^k \: \xi_l \; \rho \gamma_m \; z^{-1} \\
&&+\frac{i}{64 \pi^3} \: m^2 \: \int_x^y (\alpha^2-\alpha)
        \; R_{jk} \; \xi^j \: \xi^k \: \xi \slsh \; z^{-1} \;+\;
        {\cal{O}}(\xi^0) \;\;\; ,
\end{eqnarray*}
where $R_{jk}$ and $R$ are the (linearized) Ricci tensor and scalar 
curvature, respectively. A general difference to the formulas of 
Appendix \ref{app_A} is that $\Delta P(x,y)$ now has a stronger 
singularity on the light cone. This is a consequence of the 
$y$-derivative in (\ref{dA}). The leading singularity of $\Delta 
P(x,y)$ can be understood as describing the ``deformation'' of 
the light cone by the gravitational field in linear approximation.

We finally remark that this method works also for the 
higher-order perturbation theory as developed in \cite{F3}. It can
be used to perform the light-cone expansion of higher-order Feynman diagrams 
in the presence of a gravitational field.


\end{document}